\documentclass[12pt]{JHEP3}

\title{Center vortices as sources of Abelian dominance in pure SU(2) Yang-Mills theory}

\author{L.~E.~Oxman \\Instituto de F\'{\i}sica, Universidade Federal Fluminense,\\
Campus da Praia Vermelha, Niter\'oi, 24210-340, RJ, Brazil.\\
E-mail: oxman@if.uff.br}

\abstract{We argue that in the infrared regime of continuum Yang-Mills theory, the possibility of a mass gap in the charged sector is closely associated with the center vortex sector.

The analysis of the possible consequences of the ensembles of defects is done by showing that the description of  center vortices and monopoles is naturally unified by means of a careful treatment of Cho decomposition.

If on the one hand confinement is usually associated with monopole condensation in a compact abelian model, in this scenario, the previous decoupling of the off-diagonal degrees of freedom, for the abelian model dominate at large distances, can be understood as induced by a phase where center vortices become thick objects. 

Other important scenarios for correlated monopoles and center vortices, observed in lattice simulations,  are also accomodated in our general formulation.}

\preprint{}

\begin{document}

\section{Introduction}

In the last years, many ideas have been put forward in order to tackle the problem of confinement in pure Yang-Mills theories. A variety of scenarios such as dual superconductivity \cite{N}-\cite{hooft1}, abelian dominance \cite{cho-a,ad}, center dominance \cite{debbio3}-\cite{quandt}, the implementation of a Gribov horizon \cite{G1,G2}, and the infrared behavior of the gluon propagator \cite{AS,verS1}, have been explored.

In the mechanism of dual superconductivity, the QCD vacuum is expected to behave as
a superconductor of chromomagnetic charges, which implies the confinement of chromoelectric charges.

This nice mechanism is realized in the SU(N) Georgi-Glashow model in $(2+1)$D, where Z(N) vortices condense \cite{hooft1}, in pure compact QED in three and four dimensions, which contain monopole-like singularities \cite{polya,polya1}, and in ${\cal N}=2$ supersymmetric Yang-Mills theories \cite{sw}. 

In SU(N) pure Yang-Mills theories, one of the main problems to implement the above mentioned mechanism is the identification of the magnetic objects that could condense. For instance, monopoles can appear in the abelian projection, where a gauge fixing condition that diagonalizes a field that transforms in the adjoint representation of SU(N) is considered \cite{ap}. Monopoles also arise naturally in the representation proposed in refs. \cite{cho-a} and \cite{cho2}-\cite{Shaba}, where the gluon fields are decomposed along a general local color direction $\hat{n}$, with the advantage that no particular gauge fixing condition is invoked in this case. In these scenarios, monopole condensation has been analyzed in the lattice, by measuring different disorder parameters \cite{cp}-\cite{KLSW}, and theoretically, by studying the effective action for the magnetic background (see \cite{cho6} and references therein).

In the mechanism of abelian dominance it is conjectured that the infrared description of SU(N) Yang-Mills theories is dominated by the $N-1$ abelian gauge fields that live in the Cartan subalgebra, a phenomenon that has received support from lattice studies \cite{ad-latt}. It is generally believed that this phenomenom is a consequence of the generation of a mass gap, decoupling the charged ``off-diagonal'' degrees of freedom at large distances. In this work we will refer to abelian dominance as the generation of this gap. 

On the theoretical side, abelian dominance, in the above mentioned sense, has also been discussed in the MAG (see refs. \cite{silvio3,adom} and references therein). On the other hand, in ref. \cite{cho2}, an ``abelianized'' form for the Wilson loop in SU(2) Yang-Mills theory has been presented, not relying on any particular gauge fixing condition. However, in the this scenario, based on the consideration of Petrov-Diakonov representation for the Wilson loop integral \cite{PD} and the field decomposition along a general local color direction, the charged sector is still coupled.

The combination of abelian dominance and the identification of monopoles that condense in an effective abelian theory would provide a physical understanding of confinement in Yang-Mills theories. This is the reason why both ideas have been also explored simultaneously. For instance, in refs. \cite{CEGLP,CeaCos}, lattice simulations evidenced a dual Meissner effect together with an underlying effective Abelian theory for the SU(2) confining vacuum. Abelian dominance in the confining part of the static potential coexisting with clear signals of the dual Meissner effect, in the flux-tube profile between a quark-antiquark pair, has also been observed \cite{J5}. 

Recently, lattice studies pointed to the idea that other magnetic defects or degrees of freedom could also be relevant in the nonperturbative sector of pure QCD. The inclusion or elimination of percolating center vortices, imply quite different phases displaying confinement or deconfinement, and presence or absence of spontaneous chiral symmetry breaking (see \cite{debbio3}-\cite{quandt} and refs. therein). 

In fact, scenarios only based on either monopoles or center vortices are only partially successful to describe the behavior of the confining potential between quarks. For instance, besides linearity, they should account for N-ality dependence at asymptotic distances and Casimir scaling at intermidiate ones. One advantage in favor of the center vortex picture is that it would explain N-ality as well as it is compatible with Casimir scaling. On the other hand, the monopole picture together with abelian dominance displays some numerical success (for a review, see ref. \cite{greensite}).

One promising scenario that could accomodate the different behaviors corresponds to considering configurations where center vortices and monopoles are correlated. This correlation has received support from lattice simulations \cite{AGG}-\cite{GKPSZ} where center vortices have been observed to end at monopole worldlines.

Although, at present, we still do not know which are the more important magnetic defects to be taken into account, it has become clear that confinement has to do with the identification of the proper class of them, and the evaluation of the consequences the different associated ensembles may imply on the infrared sector of Yang-Mills theories. This is reinforced by recent theoretical results for pure Yang-Mills theory in the maximal abelian gauge (MAG) \cite{verS1}, first observed in the lattice \cite{MCM}, where the nonperturbative information is restricted to the consideration of the Gribov horizon and the possible dimension two condensates. There, the gluon and ghost propagators have been shown to be infrared suppressed. If on the one hand this behavior can be associated with the absence of gluons in asymptotic states, it also raises a question about the origin of the long range interactions responsible for quark confinement.

The aim of this article is twofold. Firstly, we will present a careful treatment of Cho decomposition \cite{cho-a}, \cite{cho2}-\cite{Shaba} when defects are present in the local color frame $n_a$, $a = 1, 2, 3$, needed to decompose the gluon fields. 
Indeed, by looking at the possible defects of the complete local color frame, we will be able to discuss not only monopoles but also center vortices in a natural and unified manner. Both objects can be associated with different types of defects the frame $\hat{n}_a$, $a=1,2,3$ can posses. This is in contrast to what happens when trying to describe defects by means of topologically nontrivial ``gauge'' transformations. While monopole-like defects can alternatively be introduced with a nontrivial transformation, as is well known, thin center vortices cannot, as the required SU(2) transformation would not be single valued along a class of closed curves, and the associated discontinuity would introduce an additional ideal vortex localized on a three-volume, besides the desired center vortex, localized on a closed two-dimensional surface (see refs. \cite{engelhardt1,reinhardt}).

Secondly, we will write the Yang-Mills partition function, including the ensemble of monopoles and center vortices coupled with the gluon fields. In this manner, we will be able to write an effective model where possible nonperturbative information associated with gluon fluctuations is parametrized in the ensembles. Up to this point, the discussion will be general and it could accomodate in the continuum anyone of the important above mentioned scenarios. At the end, we will discuss the possible relationship between abelian dominance and a sector of physical center vortices.

In section \S \ref{Cho-defects}, we obtain the possible singular terms that can appear in the field strength tensor when working with Cho decomposition and local frames containing defects. In \S \ref{m-cho} and \S \ref{center}, we show that monopoles and center vortices can be naturally unified as different topological sectors for these frames. Nonabelian transformations in terms of decomposed fields are discussed in \S \ref{Nonabe}, while SU(2) Yang-Mills theory in minimally coupled form is discussed for the maximal abelian gauge in \S \ref{YMch}. In section \S \ref{ensemble}, we derive a dual description of Yang-Mills theory that incorporates all the topological sectors. In \S \ref{emodel}, we obtain an effective model for the ensemble of defects and dual fields, showing the possible feedback of defects on the gluon sector of the theory. Finally, in section \S \ref{conc}, we present our conclusions.

\section{Cho decomposition in the presence of defects}
\label{Cho-defects}

In SU(N) Yang-Mills theory the action is given by,
\begin{equation}
{S_{YM}}=\frac{1}{2}\int d^4 x\; tr\, (F_{\mu \nu} F_{\mu \nu })
\makebox[.5in]{,}
F_{\mu \nu}=F_{\mu \nu}^{a}T^{a},
\end{equation}
where $T^a$, $a=1,..,N^2-1$ are hermitian generators of SU(N) satisfying,
\begin{equation}
\left[ T^{a},T^{b}\right]=if^{abc}T^c
\makebox[.5in]{,}
tr\, (T^a T^b)=\frac{1}{2}\delta^{ab}.
\end{equation}
As usual, the field strength tensor can be written in terms of the gauge fields $A_{\mu }^{a}$, $a=1,..,N^2-1$,
\begin{equation}
F_{\mu \nu}=(i/g)\left[D_\mu,D_\nu \right]
\label{Fcova}
\makebox[.5in]{,}
D_\mu=\partial_\mu-ig A_{\mu}^{a} T^{a},
\end{equation}
\begin{equation}
F^{a}_{\mu \nu }=\partial _{\mu }A_{\nu
}^{a}-\partial _{\nu }A_{\mu
}^{a}+g f^{abc}A_{\mu }^{b}A_{\nu }^{c}\;.
\end{equation}

For SU(2), the generators can be realized as $T^a=\tau^a/2$, $a=1,2,3$, where $\tau^a$ are the Pauli matrices, and the structure constants $f^{abc}$ are given by the Levi-Civita symbol $\epsilon^{abc}$. We will also use the notation,
\begin{equation}
\vec{F}_{\mu \nu}=\partial_\mu \vec{A}_\nu -\partial_\nu \vec{A}_\mu +g \vec{A}_\mu\times \vec{A}_\nu,
\makebox[.5in]{,}
\vec{A}_\mu=A_\mu^a\, \hat{e}_a
\makebox[.5in]{,}
\vec{F}_{\mu \nu}=F_{\mu \nu}^a\, \hat{e}_a,
\end{equation}
where $\hat{e}_a$ is the canonical basis in color space.

We take as starting point a general local frame in color space, $\hat{n}_a$, $a=1,2,3$, which can be parametrized by means of an orthogonal local transformation $R\in $ SO(3),
\begin{equation}
\hat{n}_a=R\, \hat{e}_a.
\end{equation}
This frame can be used to represent the gauge field $\vec{A}_\mu$ in terms of Cho decomposition,
\begin{equation}
\vec{A}_\mu=A^{(n)}_\mu \hat{n}-\frac{1}{g} \hat{n}\times \partial_\mu \hat{n} + \vec{X}^{(n)}_\mu
\makebox[.5in]{,}
\hat{n}.\vec{X}^{(n)}_\mu=0 ,
\label{dec}
\end{equation}
\begin{equation}
\hat{n}_a.\hat{n}_b=\delta_{ab}
\makebox[.5in]{,}
a,b=1,2,3
\makebox[.5in]{,}
\hat{n}\equiv \hat{n}_3.
\end{equation}
The restricted potential is defined as (see \cite{cho2} and references therein),
\begin{equation}
\hat{A}_\mu=A^{(n)}_\mu \hat{n}-\frac{1}{g} \hat{n}\times \partial_\mu \hat{n}.
\label{Arest}
\end{equation}
As this is already an SU(2) connection, under gauge transformations $\vec{X}^{(n)}_\mu$ transforms in the adjoint.

When dealing with a general configuration containing monopoles, the associated Dirac worldsheets, and center vortices, we will have to introduce a local frame containing defects in Euclidean spacetime. Depending on whether these configurations correspond to thin or thick objects containing a core, the parametrization above will be valid on the whole Euclidean spacetime, or only outside the core.
In the first case, we will have to deal with singular terms concentrated on the thin objects. In the second case, many 
results obtained for thin objects will also serve as an approximation for the contribution outside the cores. This comes about as the possible terms localized at the core boundaries, that may occur when manipulating the action, can be approximated by working on the whole Euclidean spacetime, including singular terms concentrated at the position of the cores.

Therefore, we consider Cho decomposition (\ref{dec}), defined on the whole Euclidean spacetime, and compute the field strength tensor,
\begin{equation}
\vec{F}_{\mu \nu}=\partial_\mu \hat{A}_\nu -\partial_\nu \hat{A}_\mu+g \hat{A}_\mu\times  \hat{A}_\nu+g \vec{X}^{(n)}_\mu\times \vec{X}^{(n)}_\nu+\hat{D}_\mu \vec{X}^{(n)}_\nu-\hat{D}_\nu \vec{X}^{(n)}_\mu,
\label{F-deco}
\end{equation}
\begin{equation}
\hat{D}_\mu \vec{X}^{(n)}_\nu=\partial_\mu \vec{X}^{(n)}_\nu+g \hat{A}_\mu\times \vec{X}^{(n)}_\nu ,
\label{Drest}
\end{equation}
keeping track of all the possible singular terms that may arise in the calculation. For the contribution associated with the restricted potential, we have,
\begin{eqnarray}
&&\partial_\mu \hat{A}_\nu -\partial_\nu \hat{A}_\mu+g \hat{A}_\mu\times  \hat{A}_\nu=F^{(n)}_{\mu \nu} \hat{n}-\frac{2}{g} \partial_\mu \hat{n}\times \partial_\nu \hat{n}
+\frac{1}{g}[\hat{n}.(\partial_\mu \hat{n} \times \partial_\nu \hat{n})]\hat{n}+\vec{L}_{\mu \nu},\nonumber \\
\label{Fcross}
\end{eqnarray}
\begin{equation}
F^{(n)}_{\mu \nu}=\partial_\mu A^{(n)}_\nu-\partial_\nu A^{(n)}_\mu
\makebox[.5in]{,}
\vec{L}_{\mu \nu}=	-\frac{1}{g} \hat{n}\times [\partial_\mu,\partial_\nu] \hat{n}=L^1_{\mu \nu}\, \hat{n}_1+L^2_{\mu \nu}\, \hat{n}_2,
\label{FL}
\end{equation}
where we have emphasized that $\vec{L}_{\mu \nu}$ is orthogonal to $\hat{n}$. This term is concentrated on two-dimensional surfaces, and could be nontrivial only for local frames containing defects in the color direction $\hat{n}$ (see the discussion in \S \ref{YMch}).
 
Now, as the second term in eq. (\ref{Fcross}) and the term $\vec{X}^{(n)}_\mu\times \vec{X}^{(n)}_\nu$ in eq. (\ref{F-deco}) are necessarily along the $\hat{n}$ direction, we can write,
\begin{equation}
\vec{F}_{\mu \nu}=(F^{(n)}_{\mu \nu}+H^{(n)}_{\mu \nu}+K_{\mu \nu}) \hat{n}+\vec{G}_{\mu \nu}+\vec{L}_{\mu \nu},
\label{FHK}
\end{equation}
\begin{equation}
\vec{G}_{\mu \nu}=\hat{D}_\mu \vec{X}^{(n)}_\nu-\hat{D}_\nu \vec{X}^{(n)}_\mu
\makebox[.3in]{,}
H^{(n)}_{\mu \nu}=-\frac{1}{g} \hat{n}.(\partial_\mu \hat{n} \times \partial_\nu \hat{n})
\makebox[.3in]{,}
K_{\mu \nu}=g\hat{n}.(\vec{X}_\mu\times \vec{X}_\nu).
\label{HK}
\end{equation}

The tensor $\vec{G}_{\mu \nu}$ have been computed in \cite{cho-a}, \cite{cho2}-\cite{cho5} and turns out to be orthogonal to $\hat{n}$, that is, it can be written in the form $\vec{G}_{\mu \nu}=G^1_{\mu \nu}\hat{n}_1+G^2_{\mu \nu}\hat{n}_2$. When singularities are present, this result remains analtered. For completness, in appendix A we include the calculation of the covariant derivative of $\vec{X}_{\mu}=X^1_{\mu}\hat{n}_1+X^2_{\mu}\hat{n}_2$, where it is obtained, 
\begin{equation}
\hat{D}_\mu \vec{X}^{(n)}_\nu=[\partial_\mu X^1_\nu-g(A^{(n)}_\mu + C^{(n)}_\mu) X^2_\nu]\hat{n}_1 +[\partial_\mu X^2_\nu+g(A^{(n)}_\mu + C^{(n)}_\mu) X^1_\nu]\hat{n}_2,
\label{DX}
\end{equation} 
\begin{equation}
C^{(n)}_\mu=-\frac{1}{g} \hat{n}_1.\partial_\mu \hat{n}_2.
\label{Cmu}
\end{equation}
On the other hand, while in refs. \cite{cho2}-\cite{cho5}, $H^{(n)}_{\mu \nu}$ has been equated with $\partial_\mu C^{(n)}_\nu -\partial_\nu C^{(n)}_\mu$, to obtain simpler ``abelianized'' expressions for the field strength tensor, when dealing with gauge fields containing defects this relationship must be revised. In fact, when defined on the whole Euclidean spacetime, both quantities differ by relevant singular terms (see \S \ref{emodel}). This difference can be obtained by noting that,
\begin{eqnarray}
\partial_\mu C^{(n)}_\nu -\partial_\nu C^{(n)}_\mu &=& -\frac{1}{g}
[\partial_\mu(\hat{n}_1.\partial_\nu \hat{n}_2)-\partial_\nu(\hat{n}_1.\partial_\mu \hat{n}_2)]\nonumber \\
&=&-\frac{1}{g}
[\partial_\mu\hat{n}_1.\partial_\nu \hat{n}_2-\partial_\nu\hat{n}_1.\partial_\mu \hat{n}_2]-\frac{1}{g}\hat{n}_1.
[\partial_\mu,\partial_\nu]\hat{n}_2.
\end{eqnarray}
Now, using $\partial_\mu\hat{n}_1.\hat{n}_1=0$, $\partial_\mu\hat{n}_2.\hat{n}_2=0$, we can write,
\begin{equation}
\partial_\mu\hat{n}_1=\alpha^1_\mu \hat{n}_2 +\beta^1_\mu \hat{n}
\makebox[.5in]{,}
\partial_\mu\hat{n}_2=\alpha^2_\mu \hat{n}_1 +\beta^2_\mu \hat{n},
\end{equation}
that is,
\begin{equation}
\partial_\mu C^{(n)}_\nu -\partial_\nu C^{(n)}_\mu=-\frac{1}{g}
[\beta^1_\mu \beta^2_\nu - \beta^1_\nu \beta^2_\mu]-\frac{1}{g}\hat{n}_1.
[\partial_\mu,\partial_\nu]\hat{n}_2.
\end{equation}
On the other hand, noting that $\hat{n}=\hat{n_1}\times \hat{n}_2$, we have,
\begin{eqnarray}
\partial_\mu\hat{n}&=&\partial_\mu \hat{n}_1 \times \hat{n}_2 + \hat{n}_1 \times \partial_\mu\hat{n}_2\nonumber \\
&=&-\beta^1_\mu \hat{n}_1 -\beta^2_\mu \hat{n}_2,
\end{eqnarray}
thus obtaining,
\begin{equation}
\hat{n}.(\partial_\mu \hat{n}\times \partial_\nu \hat{n})=
[\beta^1_\mu \beta^2_\nu-\beta^1_\nu \beta^2_\mu]=-g(\partial_\mu C^{(n)}_\nu -\partial_\nu C^{(n)}_\mu)-\hat{n}_1.
[\partial_\mu,\partial_\nu]\hat{n}_2,
\end{equation}
or, comparing with (\ref{HK}),
\begin{equation}
H^{(n)}_{\mu \nu}=\partial_\mu C^{(n)}_\nu -\partial_\nu C^{(n)}_\mu
+D^{(n)}_{\mu \nu}
\makebox[.5in]{,}
D^{(n)}_{\mu \nu}=\frac{1}{g} \hat{n}_1. [\partial_\mu,\partial_\nu]\hat{n}_2.
\label{HCD}
\end{equation}
Then, we see that the difference is nontrivial when the color directions $\hat{n}_1$, $\hat{n}_2$ contain defects.

If the frames were regular, from eqs. (\ref{FL}) and (\ref{HCD}), we would have $\vec{L}_{\mu \nu}=\vec{0}$, $\vec{D}_{\mu \nu}=\vec{0}$, and substituting in eq. (\ref{FHK}), we would obtain the abelianized ``noncompact'' simpler form given in ref. \cite{cho2}-\cite{cho5},
\begin{equation}
\vec{F}_{\mu \nu}=\left(\partial_\mu (A^{(n)}_\nu+C^{(n)}_\nu\right)-\partial_\nu (A^{(n)}_\mu+C^{(n)}_\mu)+K_{\mu \nu}) \hat{n}+\hat{D}_\mu \vec{X}^{(n)}_\nu-\hat{D}_\nu \vec{X}^{(n)}_\mu.
\label{abeFHK}
\end{equation}

In what follows, we will study eqs. (\ref{FHK})-(\ref{HCD}), exploring possible ensembles of defects and its consequences. In particular, 
to study magnetic defects, it will also be convenient to consider the associated dual expressions, defining the dual tensors using lower-case letters. For instance, the dual form of the first equation in (\ref{HCD}) reads,
\begin{equation}
h^{(n)}_{\mu \nu}=\tilde{h}^{(n)}_{\mu \nu}+d^{(n)}_{\mu \nu}
\label{hcd}
\end{equation}
\begin{equation}
h^{(n)}_{\mu \nu}=\frac{1}{2} \epsilon_{\mu \nu \rho \sigma} H^{(n)}_{\rho \sigma}
\makebox[.3in]{,}
\tilde{h}^{(n)}_{\mu \nu}=\epsilon_{\mu \nu \rho \sigma}\partial_\rho C^{(n)}_\sigma
\makebox[.3in]{,}
d^{(n)}_{\mu \nu}=\frac{1}{2} \epsilon_{\mu \nu \rho \sigma} D^{(n)}_{\rho \sigma}.
\label{h-ten}
\end{equation}

\section{Monopoles as defects of the local color frame}
\label{m-cho}

The Yang-Mills action is invariant under regular gauge transformations $S\in $ SU(N), 
\begin{equation}
\vec{A}_\mu^S.\vec{T}=S \vec{A}_\mu.\vec{T} S^{-1}+(i/g)S\partial_\mu S^{-1},
\makebox[.5in]{,}\vec{F}_{\mu \nu}^S.\vec{T}=S \vec{F}_{\mu \nu}.\vec{T} S^{-1},
\label{finite-gauge}
\end{equation}
thus implying the color currents,
\begin{equation}
J_\mu^a= g\epsilon^{abc}A_\nu^b F_{\mu \nu}^c,
\end{equation}
which at the classical level satisfy continuity equations, when computed on the equations of motion in Minkowski space. The associated conserved charges $Q^a$ satisfy a simple transformation property, when gauge transformations that assume a constant value $S_{\infty}$ at spatial infinity are considered, 
\begin{equation}
Q^a T^a \rightarrow S_{\infty} Q^a T^a S_{\infty}^{-1}.
\end{equation}

On the other hand, if on a given background $\vec{\cal A}_\mu$, a nontrivial ``gauge'' transformation is introduced,
\begin{equation}
\vec{\cal A}_\mu^{U}.\vec{T}=U \vec{\cal A}_\mu.\vec{T} U^{-1}+\frac{i}{g} U\partial_\mu U^{-1},
\label{m-gauge}
\end{equation}
where $U\in$ SU(2) is a topologically nontrivial mapping single valued along any closed loop, the Yang-Mills action changes. In particular, the field strength for $\vec{\cal A}_\mu^{U}$ is,
\begin{eqnarray}
\vec{\cal F}_{\mu \nu}^{U}.\vec{T} &=&U \vec{\cal F}_{\mu \nu}.\vec{T} U^{-1}+\frac{i}{g}U [\partial_\mu,\partial_\nu] U^{-1}.
\label{m-c}
\end{eqnarray}
As is well known, monopole-like defects can be described by considering the parametrization (\ref{m-gauge}), which  is not a simple gauge transformation. That is, the fields $\vec{\cal A}_\mu^{U}$ and $\vec{\cal A}_\mu$ are not physically equivalent, because of the second term in eq. (\ref{m-c}), which is concentrated on a two-dimensional surface where $U$ is singular. Moreover, if this parametrization were only valid outside the region where the singularites are concentrated, the field strengths would be equal on this region, but $\vec{\cal A}_\mu^{U}$ and $\vec{\cal A}_\mu$ would still be globally inequivalent. Singular terms can be accepted as long as the associated action is finite, or at least finite results are obtained after integration over trivial fluctuations around the singular background.

In order to make contact with Cho decomposition, a frame $\hat{m}_a$, $a=1,2,3$, induced by the nontrivial single valued $U$, can be introduced,
\begin{equation}
U T^a U^{-1} =\hat{m}_a.\vec{T}
\makebox[.9in]{or,}
\hat{m}_a=R(U)\, \hat{e}_a.
\label{m-frame}
\end{equation}
This frame can be parametrized in terms of Euler angles,
\begin{equation}
U= e^{-i\alpha T_3} e^{-i\beta T_2}  e^{-i\gamma T_3}
\makebox[.5in]{,}
R(U) = e^{\alpha M_3}e^{\beta M_2}e^{\gamma M_3},
\label{euler}
\end{equation}
where $M_a$ are the generators of SO(3).

Using the result obtained in ref. \cite{cho2} (in appendix B, we present an alternative derivation),
\begin{equation}
\frac{i}{g} U\partial_\mu U^{-1}=-(C^{(m)}_\mu \hat{m}+\frac{1}{g}\hat{m} \times \partial_\mu \hat{m}).\vec{T},
\label{Ugauge}
\end{equation}
\begin{equation}
C^{(m)}_\mu=-\frac{1}{g}\hat{m}_1.\partial_\mu \hat{m}_2,
\label{Cemu}
\end{equation}
we obtain,
\begin{equation}
\vec{\cal A}^{U}_\mu .\vec{T}= [({\cal A}_\mu^3-C^{(m)}_\mu) \hat{m} -\frac{1}{g}\hat{m}\times \partial_\mu \hat{m}+ {\cal A}_\mu^1 \hat{m}_1 +{\cal A}_\mu^2 \hat{m}_2 ].\vec{T}.
\end{equation}
That is, we have,
\begin{equation}
\vec{\cal A}^{U}_\mu=\vec{A}_\mu=A^{(m)}_\mu \hat{m}-\frac{1}{g} \hat{m}\times \partial_\mu \hat{m} + \vec{X}^{(m)}_\mu,
\label{m-dec}
\end{equation}
if we identify,
\begin{equation}
A^{(m)}_\mu ={\cal A}_\mu^3-C^{(m)}_\mu
\makebox[.3in]{,}
\vec{X}^{(m)}_\mu= {\cal A}_\mu^1 \hat{m}_1 + {\cal A}_\mu^2 \hat{m}_2.
\end{equation}

Both representations (\ref{m-gauge}) and (\ref{m-dec}) are equivalent when describing monopoles. However, as discussed in the next section, while (\ref{m-gauge}) cannot be extended to take into account center vortices, the consideration of Cho decomposition and an appropriate extension of the class of frames defined by the single valued $U$'s in eq. (\ref{m-frame}),  will also be useful in that case.

Now, let us consider an example, which will serve as a clue for the inclusion of center vortices in the framework of Cho decomposition. As already discussed in refs. \cite{cho2}-\cite{cho5}, monopole-like singularities in the connection can be described in terms of a defect occuring in the color direction $\hat{m}$. 
There, the consideration of $A^{(m)}_\mu=0$, $X_\mu^a=0$, $a=1,2$ and a hedgehog form $\hat{m}=\pm \hat{r}$ (color direction correlated with space direction) leads in 4D to a ``static'' Wu-Yang monopole.  In 3D Euclidean spacetime this type of configuration is called an instanton.

A straightforward calculation, using eqs. (\ref{HK}) and (\ref{h-ten}), leads to, 
\begin{equation}
g_m=\int ds_i\, h^{(m)}_{0i} = \mp \frac{4\pi}{g},
\label{m-ch}
\end{equation}
that is $\hat{m}=\hat{r}$ ($\hat{m}=-\hat{r}$) corresponds to an anti-monopole (monopole). The factor of two, with respect to the magnetic charge of a Dirac monopole, is associated with the nonabelian nature of the fields.

A configuration $\hat{m}=\hat{m}_3=\hat{r}$ can be obtained with $\alpha=\varphi$, $\beta=\theta$, where $\varphi$ and $\theta$ are the polar angles defining $\hat{r}$, and any choice for $\gamma$ can be considered, as $R\, \hat{e}_3$ is $\gamma$ independent.

For instance, we can choose $\gamma=-\varphi$. In this case, near $\theta=0$, $R\approx I$, so that the frame is not singular in the north pole on a spherical surface around the origin. On the other hand, when $\theta \approx \pi$, using $R_2(\pi)R_3(\gamma)=R_3(-\gamma)R_2(\pi)$, we get, $R \approx [R_3(\varphi)]^2 R_2(\pi)$, so that around the south pole $\hat{m}_1$, $\hat{m}_2$ is obtained from a $2\varphi$ rotation of $-\hat{e}_1$, $\hat{e}_2$, along the third axis. Then, we see that $C^{(m)}_\mu$ in eq. (\ref{Cemu}) is singular on a (Dirac) worldsheet placed on the negative $z$-axis for every (euclidean) time. 

Indeed, the calculation of $C^{(m)}_\mu$ gives (see ref. \cite{cho2}),
\begin{equation}
C^{(m)}_\mu=\frac{1}{g}(\cos \beta\, \partial_\mu \alpha +\partial_\mu \gamma),
\label{Cgamma}
\end{equation}
and the choice of $\gamma$ is associated with the position of the Dirac worldsheet. In particular, with the choice $\gamma=-\varphi$, $U$ is single valued along any closed loop, as required in eq. (\ref{m-gauge}). In this case,
\begin{equation}
C^{(m)}_\mu=\frac{1}{g}(\cos \theta -1) \partial_\mu \varphi,
\label{Cgen}
\end{equation}
and we verify that this determination is well defined on the positive $z$-axis, while the Dirac worldsheet is on the negative $z$-axis for every Euclidean time. In particular, if we go to a region close to the Dirac worldsheet ($\theta \approx \pi$), we have,
\begin{equation}
C^{(m)}_\mu \approx -\frac{2}{g} \partial_\mu \varphi.
\label{Cclose}
\end{equation}

Summarizing, while monopoles can be associated with nontrivial $\hat{m}=\hat{m}_3$ configurations, Dirac worldsheets, carrying flux $\pm 4\pi/g$, can be associated with two-di\-men\-sio\-nal defects for the associated $\hat{m}_1$, $\hat{m}_2$ components of the local frame.  In the next section we will show that general configurations, including center vortex defects, can be handled in a similar manner.

\section{Center vortices as defects of the local color frame}
\label{center}

Center vortices in SU(N) Yang-Mills theory are essentially defects in the connection such that the Wilson loop variable gains an element of the center Z(N) when the defect is linked by the Wilson loop, while it is trivial otherwise. In 4D (3D) center vortices are localized on closed two-dimensional surfaces (closed one-dimensional strings), as this type of defect is the one that can be linked by a loop. In a  thick center vortex the defining properties above are valid for Wilson loops passing up to a given minimum distance $\delta$ from the defect. 

The simpler example is, 
\begin{equation}
A^a_\mu T^a =\frac{1}{g} \partial_\mu \varphi\, \delta^{a3} T^a,
\label{thick}
\end{equation}
for $\rho > \delta$ (outside the vortex core), and a different profile for $\rho< \delta$, contributing non-trivially to the Yang-Mills action. Here, $\phi$ and $\rho$ correspond to the polar coordinates around the two-dimensional surface formed by the $z$-axis, for every Euclidean time, where the vortex is placed. For thin center vortices, $\delta \to 0$.

If a loop ${\cal C}$ passing up to a distance $\delta$ of the center vortex is considered, the Wilson loop, 
\begin{equation}
W({\cal C})=(1/2)\, tr\, P \exp (ig\oint dx_\mu A_\mu^a \tau^a/2),
\end{equation}
gives $W({\cal C})=\cos \pi=-1$, if the vortex is linked, and $W({\cal C})=1$, otherwise. 

Unlike monopoles, that can be parametrized as in eq. (\ref{m-gauge})), center vortices, even outside the vortex core, cannot be introduced by means of a nontrivial ``gauge'' transformation. This comes about because in order to generate
a profile of the type given in eq. (\ref{thick}), outside the vortex core, it would be necessary to consider an SU(2) transformation parametrized by $e^{i\varphi \, T_3}$. However, as this mapping is not single valued, we would have, 
\begin{equation}
\frac{i}{g}e^{i\varphi \, T_3}\partial_\mu e^{-i\varphi \, T_3}= \frac{1}{g} \partial_\mu \varphi\, \delta^{a3} T^a +~{\rm ideal~vortex},
\end{equation}
where the additional term (the so called ideal vortex) is localized on the three-volume where the transformation is discontinuous (see refs. \cite{engelhardt1,reinhardt}).

On the other hand, we will see that outside the cores, similarly to what happens with monopoles and their associated Dirac worldsheets, center vortices can be represented by appropriate frames containing defects. For this aim, in the case of thick objects, we will call $M$ the manifold outside the vortex cores. Its complement, $\bar{M}={\cal R}^4-M$, corresponds to (thick) closed two-dimensional sheets where the physical center vortices are localized. If the vortices were thin, we could work on the whole Euclidean spacetime, including possible singular terms in the calculations, as we have done with the Dirac worldsheets.

The possibility of matching general nontrivial configurations containing mo\-no\-poles, the associated Dirac worldsheets,  and center vortices is evidenced by parametrizing the gauge fields on $M$ in terms of Cho decomposition, based on an extended class of frames $\hat{n}_a$, obtained by introducing a $V$-sector on top of the previously considered monopole description. That is, extending $U\rightarrow VU$, we define the frame $\hat{n}_a$, $a=1,2,3$,
\begin{equation}
(VU) T^a (VU)^{-1}= V U T^a U^{-1} V^{-1} =\hat{n}_a.\vec{T}.
\label{nVU}
\end{equation}
\begin{equation}
\hat{n}_a=R(VU)\hat{e}_a=R(V)R(U)\hat{e}_a=R(V)\hat{m}_a.
\label{param-n}
\end{equation}

The defining property of $V\in$ SU(2) is that it is not single valued along any closed loop. When following a loop on $M$, we have,
\begin{equation}
	 V_f = e^{iq\pi}\, V_i,
\end{equation}
where $q=1$ or $q=0$, depending on whether the center vortex is linked or not.

For example, we can choose $V$ inducing a rotation that leaves $\hat{m}$ invariant,
\begin{equation}
	V=e^{-i\gamma_v \hat{m}.\vec{T}},
	\label{Vm}
\end{equation}
where $\gamma_v$ changes by $2\pi$ when we go around the center vortex once. Note that we can also write,
\begin{equation}
V U = e^{-i\gamma_v U T_3 U^{-1}} U= UV_3,
\makebox[.5in]{,}
V_3=e^{-i\gamma_v T_3}.
\end{equation}
In other words, the consideration of $VU$ amounts to the extension of $U$ by considering $\gamma \to \gamma +\gamma_v$ (cf. eq. (\ref{euler})). 

With the parametrization given by eqs. (\ref{param-n}) and (\ref{Vm}), we have,
\begin{equation}
\hat{n}_a =R_m(\gamma_v)\, \hat{m}_a
\makebox[.5in]{,}
R_m(\gamma_v)=e^{\gamma_v\, \hat{m}.\vec{M}},
\end{equation}
that is,
\begin{equation}
\hat{n}=\hat{m}
\makebox[.5in]{,}
h^{(n)}_{\mu \nu}=h^{(m)}_{\mu \nu},
\label{hesh}
\end{equation}
and using eq. (\ref{Cmu}),
\begin{eqnarray}
C^{(n)}_\mu &=& -\frac{1}{g}R_m\, \hat{m}_1 .
[R_m (\partial_\mu \hat{m}_2)+(\partial_\mu R_m) \hat{m}_2]=C^{(m)}_\mu +C^{(v)}_\mu,
\label{CmCv}
\end{eqnarray}
\begin{equation}
C^{(m)}_\mu=-\frac{1}{g}\hat{m}_1.\partial_\mu \hat{m}_2
\makebox[.5in]{,}
C^{(v)}_\mu=-\frac{1}{g} \hat{m}_1 . (R_m^{-1} \partial_\mu R_m)\hat{m}_2  =\frac{1}{g}\partial_\mu \gamma_v.
\label{CmCv-def}
\end{equation}
That is, $C^{(v)}_\mu$ gives the center vortex profile outside the core, with no additional ideal vortex, as $R_m$
is always single valued when we go around a closed loop. That is, although $VU$ is not in general single valued, the associated local frame $\hat{n}_a$ is, as for the adjoint representation, $R(VU)=R(-VU)$.

In other words, while monopole-like defects are associated with a nontrivial $\Pi_2$ for the space of directions $\hat{n}$, we can think of center vortices as the natural defects a frame can have, due to the nontrivial fundamental group of the adjoint representation of SU(N), needed to define the local frame to decompose the color degrees of freedom.

In this regard note that in SU(N) Yang-Mills theories, a center vortex will be associated with $V\in $ SU(N), defined outside the core, such that it changes from $V$ to $e^{i2\pi/N}\, V$ when a loop $x(u)$ linking the vortex is followed. When we go around a closed surface, following the loop, we define an open path in the fundamental representation of SU(N), $(VU)(u)$, and a nontrivial closed path $R(u)$,
$R=R(VU)$, in the adjoint representation of SU(N). This comes about as $VU$ and $e^{i2\pi/N}VU$ both define the same $R$ transformation. If this path is composed $N$ times we get a trivial map, as it is associated with a closed path in the fundamental representation of SU(N), whose first homotopy group is trivial. Then, the fundamental group of the adjoint representation of SU(N) is Z(N), which labels the possible charges of center vortices.

Summarizing, the description of monopoles, Dirac worldsheets and center vortices becomes unified, however, we have to keep in mind that, from the point of view of gauge transformations, Dirac worldsheets and center vortices  are quite different, thus opening the possibilitity to different behaviors, such as observable vortices vs. unobservable Dirac worldsheets. Note that on an open two-dimensional Dirac worldsheet the components $\hat{n}_1$ and $\hat{n}_2$ rotate twice when we go around a loop passing close and encircling this surface. However, this worldsheet is expected to be unobservable. For instance, in the example of section \S \ref{m-cho}, we can change it from the negative to the positive $z$-axis by considering an $S$ gauge transformation of the background field that corresponds to a rotation with angle $2\varphi$ (cf. eq. \ref{m-gauge}). Although this transformation is singular on the $z$-axis, it is in the trivial topological sector of SU(2), as it can be continuosly deformed onto the identity map. On the other hand, in the case of center vortices, $\hat{n}_1$ and $\hat{n}_2$ rotate once when we go around a loop linking them and, as already discussed, this cannot be associated with a nontrivial ``gauge'' transformation. 

The identification of defects in the gauge fields with topological sectors for the local frames will simplify the discussion relative to the interplay between the different degrees of freedom of the theory.

\section{Nonabelian transformations in terms of Cho variables}
\label{Nonabe}

From the discussion in the previous sections, the general ansatz on $M$ including monopole and center vortex defects 
will be given by Cho decomposition (\ref{dec}), with $\hat{n}_a$ defined by eq. (\ref{nVU}). Using the results obtained in \S \ref{Cho-defects}, the dual field strength for this decomposition is given by,
\begin{equation}
\vec{f}_{\mu \nu}=\frac{1}{2}\epsilon_{\mu \nu \rho \sigma }\vec{F}_{\rho \sigma}=(f^{(n)}_{\mu \nu}+h^{(n)}_{\mu \nu}+k_{\mu \nu})\hat{n}+ (g_1^{\mu \nu}+l_1^{\mu \nu})\hat{n}_1 +(g_2^{\mu \nu}+l_2^{\mu \nu}) \hat{n}_2,
\label{dualsing}
\end{equation}
where the tensors in lower-case are the dual ones associated with $F^{(n)}_{\mu \nu}$, $H^{(n)}_{\mu \nu}$, $K_{\mu \nu}$, $G_i^{\mu \nu}$ and $L_i^{\mu \nu}$ in eqs. (\ref{FL})-(\ref{DX}). 
That is,
\begin{equation}
f^{(n)}_{\mu \nu}=\frac{1}{2}\epsilon_{\mu \nu \rho \sigma }F^{(n)}_{\rho \sigma}
\makebox[.3in]{,}
{l}_1^{\mu \nu}\hat{n}_1+{l}_2^{\mu \nu}\hat{n}_2=\frac{1}{2}\epsilon_{\mu \nu \rho \sigma }\vec{L}^{\mu \nu}
\makebox[.3in]{,}
k_{\mu \nu}=g \epsilon_{\mu \nu \rho \sigma} X_\rho^1 X_\sigma^2,
\label{felek}
\end{equation}
\begin{eqnarray}
g_1^{\mu \nu}&=&\epsilon^{\mu \nu \rho \sigma} [\partial_\rho X^1_\sigma-g(A^{(n)}_\rho+C^{(n)}_\rho)X^2_\sigma],\nonumber \\
g_2^{\mu \nu}&=&\epsilon^{\mu \nu \rho \sigma} [\partial_\rho X^2_\sigma+g(A^{(n)}_\rho+C^{(n)}_\rho)X^1_\sigma],
\label{ges}
\end{eqnarray}
see also eq. (\ref{h-ten}). Then, for thick vortices, the Yang-Mills action can be written as, 
\begin{equation}
S_{YM}=S_{\bar{M}}+S_M,
\end{equation}
\begin{equation}
S_M=\frac{1}{4} \int_M d^4x\,  [(f^{(n)}_{\mu \nu} +h^{(n)}_{\mu \nu}+k_{\mu \nu})^2 + 
(g_1^{\mu \nu}+l_1^{\mu \nu})^2 +(g_2^{\mu \nu}+l_2^{\mu \nu})^2].
\label{YM-fhk}
\end{equation}
The term $S_{\bar{M}}$  gives the contribution coming from the vortex cores. As already discussed, when vortices are thin, this term is absent and the replacement $M\rightarrow {\cal R}^4$ must be considered in eq. (\ref{YM-fhk}). 

Although Cho decomposition assumes an abelianized form, specially when we look at a local color frame containing no defects, and the fields $A^{(n)}_\mu$, $\vec{X}^{(n)}_\mu$, (see eq. \ref{abeFHK}), it is important to keep in mind that the full nonabelian degrees must be represented in this formulation, as it is equivalent to the underlying Yang-Mills theory.

In section \S \ref{ensemble}, we will take $A^{(n)}_\mu$, $\vec{X}^{(n)}_\mu$ as independent variables, so that it is important to discuss how nonabelian gauge transformations can be translated into ``abelianized'' language. Let us consider a gauge transformation of the gauge field $\vec{A}_\mu$ given in eq. (\ref{dec}), decomposed in terms of a general frame $\hat{n}_a$, possibly containing monopole and center vortex defects,
\begin{equation}
\vec{A}^S_\mu.\vec{T}= S(A^{(n)}_\mu \hat{n}-\frac{1}{g} \hat{n}\times \partial_\mu \hat{n} + \vec{X}^{(n)}_\mu).\vec{T}S^{-1}+\frac{i}{g}S\partial_\mu S^{-1}
\end{equation}
Because of eqs. (\ref{dec}), (\ref{hesh}) and (\ref{U-trans}), we can also write,
\begin{eqnarray}
\vec{A}^S_\mu.\vec{T}&=&S[(A^{(n)}_\mu+C^{(m)}_\mu)\hat{n}+ \vec{X}^{(n)}_\mu].\vec{T}S^{-1}+\frac{i}{g}S  (U\partial_\mu U^{-1})S^{-1}+\frac{i}{g}S\partial_\mu S^{-1}\nonumber \\
&=&S[(A^{(n)}_\mu+C^{(m)}_\mu)\hat{n}+ X^1_\mu\, \hat{n}_1+ X^2_\mu\, \hat{n}_2].\vec{T}S^{-1}+\frac{i}{g}(SU)\partial_\mu (SU)^{-1},
\end{eqnarray}
and using again eq. (\ref{U-trans}), we get,
\begin{eqnarray}
\vec{A}^S_\mu.\vec{T}&=&[(A^{(n)}_\mu+C^{(m)}_\mu)\hat{n}'+ X^1_\mu\, \hat{n}'_1+ X^2_\mu\, \hat{n}'_2].\vec{T}
-(C^{(m')}_\mu \hat{m}'+\frac{1}{g}\hat{m}' \times \partial_\mu \hat{m}').\vec{T},\nonumber \\
\end{eqnarray}
\begin{equation}
\hat{n}'_a .\vec{T} = S \hat{n}_a .\vec{T} S^{-1}
\makebox[.5in]{,}
\hat{m}'_a .\vec{T} = S \hat{m}_a .\vec{T} S^{-1}.
\label{mprima}
\end{equation}
Note that according to appendix B, the color directions $\hat{m}'_a$, $a=1,2$, are the ones necessary to compute $C^{(m')}_\mu$, while $\hat{m}'_3\equiv \hat{m}'$. From eq. (\ref{mprima}), and using $\hat{n}_a .\vec{T}=V \hat{m}_a.\vec{T} V^{-1}$ (see section \S \ref{center}), we also obtain,
\begin{eqnarray}
\hat{n}'_a .\vec{T}=V' \hat{m}'_a.\vec{T} V'^{-1}
\makebox[.5in]{,}
V'=SVS^{-1}=e^{-i\gamma_v S\hat{m}.\vec{T}S^{-1}}=e^{-i\gamma_v \hat{m}'.\vec{T}}.
\label{sim-linha}
\end{eqnarray}
That is, $\hat{n}'=\hat{m}'$, and the gauge transformed field is,
\begin{equation}
\vec{A}^S_\mu=A^{(n')}_\mu \hat{n}'-\frac{1}{g} \hat{n}'\times \partial_\mu \hat{n}' + X^1_\mu\, \hat{n}'_1+ X^2_\mu\, \hat{n}'_2,
\end{equation}
\begin{equation}
A^{(n')}_\mu=A^{(n)}_\mu+ C^{(m)}_\mu - C^{(m')}_\mu
\makebox[.5in]{,}
\hat{n}'_a = R(S)\, \hat{n}_a.
\label{nonabe-abe}
\end{equation}
Now, using eqs. (\ref{YM-fhk}), (\ref{hcd}) and (\ref{hesh}) we see that the effect of a nonabelian gauge transformation in the Yang-Mills action is,
\begin{equation}
S'_{YM}=S'_{\bar{M}}+S'_M,
\end{equation}
\begin{eqnarray}
S'_{M}&=&\frac{1}{4}  \int_M d^4x\, [(f^{(n)}_{\mu \nu}+\tilde{h}^{(m)}_{\mu \nu}-
\tilde{h}^{(m')}_{\mu \nu}+h^{(n')}_{\mu \nu}+k_{\mu \nu})^2 + ({g'}_1^{\mu \nu}+{l'}_1^{\mu \nu})^2+ ({g'}_2^{\mu \nu}+{l'}_2^{\mu \nu})^2]\nonumber \\
&=&\frac{1}{4}  \int_M d^4x\, [(f^{(n)}_{\mu \nu}+h^{(n)}_{\mu \nu}+d^{(m')}_{\mu \nu}-d^{(m)}_{\mu \nu}+k_{\mu \nu})^2 + (g_1^{\mu \nu}+{l'}_1^{\mu \nu})^2+ (g_2^{\mu \nu}+{l'}_2^{\mu \nu})^2],\nonumber \\
\label{YM-linha}
\end{eqnarray}
where in the expressions for ${g'}_a^{\mu \nu}$, $a=1,2$, depending on $A^{(n')}_\mu+C^{(n')}_\mu$ (cf. eq. (\ref{ges})), we used eqs. (\ref{CmCv}) and (\ref{CmCv-def}), and a similar one derived from eq. (\ref{sim-linha}),
\begin{eqnarray}
C^{(n')}_\mu = C^{(m')}_\mu +\frac{1}{g} \partial_\mu \gamma_v,
\end{eqnarray}
to obtain the invariance,
\begin{equation}
A^{(n')}_\mu+C^{(n')}_\mu=A^{(n)}_\mu+ C^{(m)}_\mu - C^{(m')}_\mu+C^{(n')}_\mu=A^{(n)}_\mu+ C^{(n)}_\mu.
\label{AmasC}
\end{equation}

Of course, when the frames are regular, it is evident that the action is gauge invariant. 
In turn, if gauge symmetry is preserved, the only additional terms, when comparing $S_{M}$ and $S'_{M}$ (cf. eqs. (\ref{YM-fhk}) and (\ref{YM-linha})), must correspond to a change of the (unobservable) Dirac worldsheets (see section \S \ref{ensemble}). 

\section{Yang-Mills charged fields in minimally coupled form}
\label{YMch}

If in the previous section, we consider a gauge transformation $S$, representing a frame rotation with regular angle $\chi$, leaving the local color direction $\hat{n}$ invariant, eq. (\ref{nonabe-abe}) gives, 
\begin{equation}
A^{(n')}_\mu=A^{(n)}_\mu- \frac{1}{g}\partial_\mu \chi,
\end{equation}
$d^{(m')}_{\mu \nu}-d^{(m)}_{\mu \nu}=0$, and ${l'}_a^{\mu \nu}=l_a^{\mu \nu}$. If the defects are chosen such that the singular terms $l_a^{\mu \nu}$, $a=1,2$ are nulified, a rotation of the basis elements $\hat{n}_1$, $\hat{n}_2$ can be translated to a U(1) rotation of the components $X_\mu^1$, $X_\mu^2$ (leaving the basis elements fixed),
\begin{equation}
A^{(n)}_\mu \rightarrow A^{(n)}_\mu -\frac{1}{g}\partial_\mu \chi
\makebox[.5in]{,}
\Phi_\mu \rightarrow e^{i\chi} \Phi_\mu,
\label{Uum}
\end{equation}
\begin{equation}
\Phi_\mu=\frac{1}{\sqrt{2}}(X^1_\mu+iX^2_\mu).
\label{Fi}
\end{equation}
Then, in order to associate the fields $\Phi_\mu$ with a well defined charged sector, we will take $\hat{n}$ as the direction in color space associated with those defects localized on closed strings (monopoles), while the defects in $\hat{n}_a$, $a=1,2$ will be associated with Dirac worldsheets and center vortices. In this case, as the $\hat{n}$ sector does not contain defects localized on two-dimensional worldsheets, the singular terms $l_a^{\mu \nu}$, ${l'}_a^{\mu \nu}$, $a=1,2$ vanish. 

Note that this ensemble of defects is invariant under regular gauge transformations. In addition, if a singular gauge transformation $S$ along the direction $\hat{n}$, living in the trivial topological sector of $SU(2)$, is considered, the term $d^{(m)}_{\mu \nu}-d^{(m')}_{\mu \nu}$ in eq. (\ref{YM-linha}) will in general be nonzero, representing at most a trivial flux $4\pi/g$, concentrated on a closed two-dimensional worldsheet (see the example in section \S \ref{m-cho}).

Now, in order to emphazise the abelian aspects of the decomposition, let us introduce the first order formalism, and define the MAG gauge condition.  

\subsection{Yang-Mills action}

The Yang-Mills action can be written as,
\begin{eqnarray}
S_{YM}&=&S_{\bar{M}}+\int_M d^4x\, \frac{1}{4} [(f^{(n)}_{\mu \nu} +h^{(n)}_{\mu \nu}+k_{\mu \nu})^2 + 
g_1^{\mu \nu} g_1^{\mu \nu}+ g_2^{\mu \nu} g_2^{\mu \nu}]\nonumber \\
&=&S_{\bar{M}}+\int_M d^4x\, \left[ \frac{1}{4} (f^{(n)}_{\mu \nu} +h^{(n)}_{\mu \nu}+k_{\mu \nu})^2 + \frac{1}{2} \bar{g}^{\mu \nu} g^{\mu \nu}\right],
\label{SM}
\end{eqnarray}
where we have defined,
\begin{equation}
g^{\mu \nu}=\frac{1}{\sqrt{2}}(g_1^{\mu \nu}+ig_2^{\mu \nu})=\epsilon^{\mu \nu \rho \sigma}[\partial_\rho+ig(A^{(n)}_\rho+C^{(n)}_\rho)]\Phi_\sigma,
\end{equation}
\begin{equation}
k_{\mu \nu}=\frac{g}{2i}\epsilon_{\mu \nu \rho \sigma}(\bar{\Phi}_\rho \Phi_\sigma-\Phi_\rho \bar{\Phi}_\sigma),
\end{equation}
with $\Phi_\mu$ given by eq. (\ref{Fi}). 

Introducing real and complex lagrange multipliers, $\lambda_{\mu \nu}$ and $\Lambda_{\mu \nu}$,
\begin{eqnarray}
\lefteqn{S_{M}=}\nonumber \\
&&=S_c+\int_M d^4x\, \left[\frac{1}{4}\lambda_{\mu \nu} \lambda_{\mu \nu} -\frac{i}{2}
\lambda_{\mu \nu}(f^{(n)}_{\mu \nu}+h^{(n)}_{\mu \nu}+k_{\mu \nu})+i J^\mu (A^{(n)}_\mu+C^{(n)}_\mu) \right],\nonumber \\
\end{eqnarray}
\begin{equation}
S_c=\int_M d^4x\, \left[\frac{1}{2}\bar{\Lambda}^{\mu \nu} \Lambda^{\mu \nu}-\frac{i}{2} (\bar{\Lambda}^{\mu \nu} 
\epsilon^{\mu \nu \rho \sigma}\partial_\rho \Phi_\sigma + {\Lambda}^{\mu \nu} 
\epsilon^{\mu \nu \rho \sigma}\partial_\rho \bar{\Phi}_\sigma)\right],
\end{equation}
Here, we can read the action for charged fields minimally coupled to the U(1) color current,
\begin{equation}
J^\mu = -\frac{i}{2}\, g \epsilon^{\mu \nu \rho \sigma} \bar{\Lambda}_{\nu \rho}\Phi_\sigma + \frac{i}{2}\, g \epsilon^{\mu \nu \rho \sigma} {\Lambda}_{\nu \rho}\bar{\Phi}_\sigma.
\label{Jlambda}
\end{equation}
Note that the U(1) symmetry now reads,
\begin{equation}
A^{(n)}_\mu \rightarrow A^{(n)}_\mu -\partial_\mu \chi
\makebox[.5in]{,}
\Phi_\mu \rightarrow e^{ig\chi} \Phi_\mu,
\makebox[.5in]{,}
\Lambda_{\mu \nu}\rightarrow e^{ig\chi} \Lambda_{\mu \nu}.
\end{equation}

\subsection{Gauge fixing}

With regard to gauge fixing, we will adopt, for the charged part on M (for a discussion in the context of Cho decomposition, see ref. \cite{kondo6}),
\begin{equation}
\hat{D}_\mu \vec{X}^{(n)}_\mu =0,
\label{MAG}
\end{equation}
while for the diagonal fields, we will consider,
\begin{equation}
\partial_\mu (A^{(n)}_\mu+C^{(n)}_\mu)=0.
\label{lorentz}
\end{equation}
These conditions can be imposed by means of lagrange multipliers $\vec{b}=b_1 \hat{n}_1+b_2 \hat{n}_2$, $\beta$, for the gauge fixings (\ref{MAG}) and (\ref{lorentz}), respectively, and including in the path integral the factor,
\begin{equation}
e^{i\int_M d^4x\, [\beta \partial_\mu (A^{(n)}_\mu+C^{(n)}_\mu)+\vec{b}. \hat{D}_\mu \vec{X}^{(n)}_\mu]}.
\label{gf-factor}
\end{equation}
In addition, we will have a Fadeev-Popov determinant, exponentiated by means of the associated ghost fields $\vec{c}=c_1 \hat{n}_1+c_2 \hat{n}_2$ and $c$. The action for the ghosts contains a term quadratic in $\hat{D}_\mu$,
\begin{equation}
\int_M d^4x\, \vec{c}^{\;\ast}. \hat{D}_\mu \hat{D}_\mu \vec{c},
\end{equation}
where $\vec{c}^{\;\ast}=\bar{c}_1 \hat{n}_1+\bar{c}_2 \hat{n}_2$.
This term can be linearized by  introducing additional auxiliary fields $\vec{a}^\mu=a^\mu_1 \hat{n}_1+a^\mu_2 \hat{n}_2$, and a factor of the form,
\begin{equation}
e^{i\int_M d^4x\, ( \vec{a}^{\;\ast}_\mu .\hat{D}_\mu \vec{c}+ c.c)}\, e^{-\int_M d^4x\, \vec{a}^{\;\ast}_\mu .\vec{a}_\mu}.
\label{factor-ghost}
\end{equation}
As shown in ref. \cite{silvio3,silvio1}, the renormalization procedure typically introduces additional quartic ghost terms and other terms coupling the ghosts and the charged fields $\vec{X}_\mu$, containing up to linear terms in $\hat{D}_\mu$.

Using eq. (\ref{DX}), we can rewrite eq. (\ref{MAG}) as,
\begin{equation}
{\cal D}_\mu \Phi_\mu=0
\makebox[.5in]{,}
\bar{{\cal D}}_\mu \bar{\Phi}_\mu=0,
\end{equation}
where we have defined,
\begin{equation}
{\cal D}_\mu =\partial_\mu + ig(A^{(n)}_\mu+C^{(n)}_\mu)
\makebox[.5in]{,}
\bar{{\cal D}}_\mu =\partial_\mu - ig(A^{(n)}_\mu+C^{(n)}_\mu),
\end{equation}
and the factor (\ref{gf-factor}) results,
\begin{equation}
e^{i\int_M d^4x\, \left[\beta \partial_\mu (A^{(n)}_\mu+C^{(n)}_\mu)+\bar{b}\, {\cal D}_\mu \Phi_\mu +
b\, \bar{{\cal D}}_\mu \bar{\Phi}_\mu \right]}
\makebox[.5in]{,}
b=\frac{1}{\sqrt{2}}(b_1 +i b_2).
\end{equation}
Now, using a formula like (\ref{DX}), for $\hat{D}_\mu \vec{c}$,
\begin{equation}
\hat{D}_\mu \vec{c} =[\partial_\mu c_1 -g(A^{(n)}_\mu+C^{(n)}_\mu) c_2]\hat{n}_1
+[\partial_\mu c_2+g(A^{(n)}_\mu+C^{(n)}_\mu)c_1]\hat{n}_2,
\end{equation}
the factor (\ref{factor-ghost}) takes the form,
\begin{equation}
e^{i\int_M d^4x\, \left[ \bar{a}_1^\mu [\partial_\mu c_1 -g(A^{(n)}_\mu+C^{(n)}_\mu) c_2]
+\bar{a}_2^\mu [\partial_\mu c_2+g(A^{(n)}_\mu+C^{(n)}_\mu)c_1]+ c.c\right]}\, e^{-\int_M d^4x\, \vec{a}^{\;\ast}_\mu .\vec{a}_\mu}.
\end{equation}
The important point is that the gauge fixing part of the measure depends on the combination $A^{(n)}_\mu+C^{(n)}_\mu$, and can be written as,
\begin{equation}
F_{gf}=F^{c}_{gf}\,  e^{-i\int_M d^4x\, (A^{(n)}_\mu+C^{(n)}_\mu)K^\mu} ,
\end{equation}
where $F^{c}_{gf}$ is independent of $A^{(n)}_\mu$, collecting all the other factors and the integration measure
for lagrange multipliers, ghosts and auxiliary fields, while $K^\mu=\partial^\mu \beta + ...$, besides these fields,  also depends on $\Phi_\mu$.

\section{General ensembles of defects}
\label{ensemble}

As we discussed in \S \ref{m-cho} and \S \ref{center}, one advantage of using Cho decomposition to pa\-ra\-metrize the gauge fields is that monopoles and center vortices can be represented on the same footing, by means of a general local color frame containing defects. On the other hand, it is important to remark that while monopole defects can alternatively be associated with topologically nontrivial ``gauge'' transformations, thin center vortices cannot, as a gauge transformation using an SU(2) transformation multivalued along a closed loop would also introduce an ideal vortex concentrated on a three-volume. 

This opens the possibility of two different behaviors. While open Dirac worldsheets carrying flux $4\pi/g$ remain unobservable, if gauge symmetry is preserved, no symmetry is present to protect thin center vortices from a destabilization into physical thick center vortices.

The use of monopoles and center vortices as a reasonable phase on top of which gluon fluctuations can be included depends on their stability. In turn, this stability can be studied by means of the path integration over gluon fields in the given background, and analyzing if $\log Z_{YM}$ aquires a real part due to effective bound states of the charged fields. In that case $|Z_{YM}|^2$, the probability to persist in the fundamental state of the theory, in a given background, would be less than one, thus signaling instability.

In ref. \cite{cho5}, the stability of different magnetic configurations have been analyzed to one loop in the gluon fields.
Although some controversy exists with regard to the stability of center vortices in this approximation \cite{bordag}, lattice simulations \cite{debbio3} point to the idea that they become observable physical objects, with a thickness of the order of 1fm. That is, the thin center vortices are expected to be stabilized by generating a finite radius (see also the discussion in refs. \cite{engelhardt1} and \cite{diakonov-center}).

On the other hand, monopoles lead to an unstable phase, which is expected to be stabilized by including monopole-monopole interactions (for a review see \cite{antonov} and refs. therein).

In general, for a given gauge field $A^ a_\mu$, $a=1,2,3$, many different local frames $\hat{n}_a$ can be introduced to decompose it. 
In refs. \cite{kondo3,kondo7}, Cho variables have been incorporated by including, in the partition function for Yang-Mills theory, an identity written as an integral over local color directions $\hat{n}$, satisfying $\hat{n}.\hat{n}=1$, and then showing that the Jacobian of the transformation,
\[
\vec{A}_\mu, \hat{n} \rightarrow A^{(n)}_\mu, X^{(n)}_\mu, \hat{n} ,
\]
is trivial. The integration over $\hat{n}$ and the constraint can be represented as an integration over single valued $U$ transformations, defining $\hat{n}=\hat{m}=\hat{m}_3$ through eq. (\ref{euler}). In our case, this integration includes a summation over different classes of $U$'s, implying different locations for the monopole singularities. with a given prescription for the associated Dirac worldsheets, to be discussed below.
In addition, the center vortex sector will be given by an additional summation over classes of $V$'s, producing distributions of center vortices, on top of the monopole configurations. 

Then, according to the previous discussions, gauge fields with defects will be taken into account by considering $A_\mu^{(n)}$, $X^1_\mu$, $X^2_\mu$ as regular fields, and using the parametrization summarized by eqs. (\ref{dec}) and (\ref{nVU}), defining a local color frame $\hat{n}_a$, $a=1,2,3$, containing monopoles, the associated Dirac worldsheets, and center vortex defects. In particular, according to the discussion in section \S \ref{YMch}, the Yang-Mills partition function can be represented as,
\begin{eqnarray}
Z_{YM} 
& =& \int [{\cal D}V][{\cal D}U][{\cal D}A^{(n)}][{\cal D}\Phi] F_{gf}\, e^{-S_{YM}} \nonumber \\
&= &\int [{\cal D}\lambda][{\cal D}V][{\cal D}U][{\cal D}A^{(n)}][{\cal D}\Phi]
[{\cal D}\Lambda] F^{c}_{gf}\, e^{-S_{\bar{M}}-S_c-\int_M d^4x\, \frac{1}{4}\lambda_{\mu \nu} \lambda_{\mu \nu}}\times\nonumber \\
&&\times e^{i\int_M d^4x\, [\frac{1}{2}\lambda_{\mu \nu}(f^{(n)}_{\mu \nu}+h^{(n)}_{\mu \nu}+k_{\mu \nu})- J_c^\mu (A^{(n)}_\mu+C^{(n)}_\mu)]},\nonumber \\
\label{ZYM-exa}
\end{eqnarray}
\begin{equation}
	J_c^\mu=J^\mu +K^\mu,
\end{equation}
where appropriate boundary conditions are implicit in the path integral measure (finite temperature periodic conditions, etc.). 

It will also be convenient to consider the following parametrization,
\begin{equation}
\lambda_{\mu \nu}=\partial_\mu \phi_\nu-\partial_\nu \phi_\mu +B_{\mu \nu},
\label{lambda}
\end{equation}
with,
\begin{equation}
\partial_\mu \phi_\mu=0  \makebox[.5in]{,}  \partial_\nu B_{\mu \nu}=0,
\label{g-fixing}
\end{equation}
also replacing in eq. (\ref{ZYM-exa}),
\begin{equation}
[{\cal D}\lambda]\to [{\cal D}B][{\cal D}\phi] F^{B}_{gf}F^{\phi}_{gf},
\label{delambda}
\end{equation}
where $F^{B}_{gf}$ is the part of the measure fixing the condition $\partial_\nu B_{\mu \nu}=0$,
\begin{equation}
F^{B}_{gf}=[{\cal D}\xi_\mu] e^{i\int d^4x\, \xi_\mu \partial_\nu B_{\mu \nu}},
\label{xi-fixing}
\end{equation}
and $F^{\phi}_{gf}$ is the part of the measure fixing the gauge $\partial_\mu \phi_\mu=0$,
\begin{equation}
F^{\phi}_{gf}=[{\cal D}\xi] e^{i\int d^4x\, \xi \partial_\mu \phi_\mu}.
\end{equation}

We recall, that in the case of thin center vortices, we have to consider $S_{\bar{M}}=0$ and $M\rightarrow {\cal R}^4$. On the other hand, for thick center vortices,
the path integral measure in eq. (\ref{ZYM-exa}) must also include a gauge fixed path integral over the fields on $\bar{M}$, inside the vortex cores. In this case, at the boundary of $\bar{M}$, the gauge fields and the gauge fixing conditions must be matched with those given in eqs. (\ref{dec}) and (\ref{MAG}), respectively. 

Now, we can integrate by parts the term containing $f^{(n)}_{\mu \nu}$ and note that because of the $A^{(n)}_\mu$ path integration, a constraint is implicit here for the fields defined on $M$, 
\begin{equation}
J_c^\mu=\frac{1}{2}\epsilon_{\mu \nu \rho \sigma}\partial_\nu \lambda_{\rho \sigma}=\frac{1}{2}\epsilon_{\mu \nu \rho \sigma}\partial_\nu B_{\rho \sigma},
\label{j-map}
\end{equation}
which implies,
\begin{equation}
\partial_\mu J_c^\mu=0.
\end{equation}
That is, we can consider the replacement,
\begin{equation} 
\int_M d^4x\,J_c^\mu C^{(n)}_\mu\to \int_M d^4x\, \frac{1}{2}\epsilon_{\mu \nu \rho \sigma}\partial_\nu \lambda_{\rho \sigma} C^{(n)}_\mu.  
\end{equation}
Again, integrating by parts and using eqs. (\ref{hcd}) and (\ref{h-ten}) we arrive at,
\begin{eqnarray}
Z_{YM} 
&= &\int [{\cal D}\lambda][{\cal D}V][{\cal D}U][{\cal D}A^{(n)}][{\cal D}\Phi]
[{\cal D}\Lambda] F^{c}_{gf}\, e^{-S_{\bar{M}}-S_c-\int_M d^4x\, \frac{1}{4}\lambda_{\mu \nu} \lambda_{\mu \nu}}\times\nonumber \\
&&\times e^{i\int_M d^4x\, \{ (\frac{1}{2}\epsilon_{\mu \nu \rho \sigma} \partial_\nu \lambda_{\rho \sigma}-
J^c_\mu ) A^{(n)}_\mu +\frac{1}{2}\lambda_{\mu \nu}(d^{(n)}_{\mu \nu}+k_{\mu \nu})+\partial_\mu[\frac{1}{2}(A^{(n)}_\sigma+C^{(n)}_\sigma)\epsilon_{\mu \nu \rho \sigma} \lambda_{\nu \rho}]\}},\nonumber \\
\label{ZYMb}
\end{eqnarray}
where $[{\cal D}\lambda]$ is given by eq. (\ref{delambda}).

\subsection{Singular terms}

Because of eqs. (\ref{hesh})-(\ref{CmCv-def}), the contributions to $d^{(n)}_{\mu \nu}$, associated with the monopole and vortex ensembles, become separated,
\begin{equation}
d_{\mu\nu}^{(n)}=h^{(n)}_{\mu \nu}-\tilde{h}^{(n)}_{\mu \nu}=
d^{(m)}_{\mu\nu} +d^{(v)}_{\mu\nu},
\label{mv-sep}
\end{equation}
where,
\begin{equation} 
d^{(m)}_{\mu\nu}=h^{(m)}_{\mu \nu}-\tilde{h}^{(m)}_{\mu \nu},
\end{equation}
\begin{equation}
d^{(v)}_{\mu\nu}=-\tilde{h}^{(v)}_{\mu \nu}=-\frac{1}{g}\epsilon_{\mu \nu \rho \sigma} \partial_\rho \partial_\sigma \gamma_v.
\label{dv}
\end{equation} 
Let us first discuss $d^{(m)}_{\mu\nu}$ by considering the example in section \S \ref{m-cho}, where $\hat{m}_3=\hat{r}$.
The difference between $h^{(m)}_{\mu \nu}$ and $\tilde{h}^{(m)}_{\mu \nu}=\epsilon_{\mu \nu \rho \sigma}\partial_\rho C^{(n)}_\sigma$ is associated with singularities in the behavior of $C^{(m)}_\mu$. Close to the Dirac worldsheet, 
$C^{(m)}_\mu=-\frac{2}{g} \partial_\mu \varphi+~{\rm regular~term}$ (see eq. (\ref{Cclose})), and because of the singularity of $\varphi$ on the $z$-axis,
\begin{equation}
\tilde{h}^{(m)}_{0i}=h^{(m)}_{0i}-\frac{4\pi}{g} \delta^{(2)}(x_1,x_2) \theta(-x_3)\delta_{i3},
\end{equation}
\begin{equation}
d^{(m)}_{0i}= \frac{4\pi}{g} \delta^{(2)}(x_1,x_2) \theta(-x_3)\delta_{i3}.
\end{equation}
That is, in the whole Euclidean spacetime, the only difference between $h^{(m)}_{0i}$ and $\tilde{h}^{(m)}_{0i}$ is that the ``field lines'' of the second are closed.

Of course, in 4D the monopole above corresponds to an infinite string-like defect, placed at $\vec{x}=0$,
at every time $x_0$, which defines the border of the open two-dimensional Dirac worldsheet. In general, the relevant monopole string-like defects, having infrared finite euclidean action, must be in fact closed strings. 
In this case, the difference $d^{(m)}_{\mu \nu}$ contains an additional contribution concentrated on an open two-dimensional Dirac worldsheet (the string-like monopoles are at the border) such that,
$\partial_\nu \tilde{h}^{(m)}_{\mu \nu}=0$, everywhere. As a consequence,
\begin{eqnarray}
d_{\mu\nu}^{(m)}=h^{(m)}_{\mu\nu} -\tilde{h}^{(m)}_{\mu\nu} &=&\frac{4\pi}{g} \int d\sigma_1 d\sigma_2\, 
\left(\frac{\partial x_\mu}{\partial \sigma_1}\frac{\partial x_\nu}{\partial \sigma_2}-
\frac{\partial x_\mu}{\partial \sigma_2}\frac{\partial x_\nu}{\partial \sigma_1}\right) \delta^{(4)}(x-x(\sigma_1,\sigma_2))\nonumber \\
&=&\frac{4\pi}{g} \int d^2 \sigma_{\mu \nu}\, \delta^{(4)}(x-x(\sigma_1,\sigma_2)).
\end{eqnarray}
Here, $x(\sigma_1,\sigma_2)$ is the Dirac worldsheet for a monopole anti-monopole pair.
 
Considering that $\sigma_2 \in [0,1]$ corresponds to the periodic direction, 
\[
x_\mu(\sigma_1,0)=x_\mu(\sigma_1,1),
\]
while $\sigma_1 \in [0,1]$ corresponds to the open direction, we obtain,
\begin{eqnarray}
\partial_\nu d_{\mu \nu}^{(m)}&=&
\frac{4\pi}{g} \int d\sigma_1 d\sigma_2\, 
\left(\frac{\partial}{\partial \sigma_1}\left(\frac{\partial x_\mu}{\partial \sigma_2}\delta^{(4)}\right)-\frac{\partial}{\partial \sigma_2}\left(\frac{\partial x_\mu}{\partial \sigma_1}\delta^{(4)}\right)\right) \nonumber \\ 
&=&\frac{4\pi}{g} \int d\sigma_1 d\sigma_2\, 
\frac{\partial}{\partial \sigma_1}\left(\frac{\partial x_\mu}{\partial \sigma_2}\delta^{(4)}\right)\nonumber \\
&=&\frac{4\pi}{g} \int d\sigma_2\, \left(\left.\frac{\partial x_\mu}{\partial \sigma_2}\right|_{\sigma_2=1}\delta^{(4)}(x-x(1,\sigma_2))-\left.\frac{\partial x_\mu}{\partial \sigma_2}\right|_{\sigma_2=0}\delta^{(4)}(x-x(0,\sigma_2))\right)\nonumber \\
&=&\frac{4\pi}{g} \left( \oint_{C^+} dy_\mu\, \delta^{(4)}(x-y)- \oint_{C^-} dy_\mu\, \delta^{(4)}(x-y) \right),
\label{divd}
\end{eqnarray}
where $C^+$ ($C^-$) is the loop where the monopole (anti-monopole) is localized.

Similarly, as $\gamma_v$ is singular on a closed two-dimensional surface, changing by $2\pi$ when we go around it, when eq. (\ref{dv}) is extended to the whole Euclidean spacetime, $d^{(v)}_{\mu\nu}$ is nontrivial, and for a general center vortex, 
\begin{eqnarray}
d_{\mu \nu}^{(v)}&=&
\frac{2\pi}{g} \int d\sigma_1 d\sigma_2\, 
\left(\frac{\partial x_\mu}{\partial \sigma_1}
\frac{\partial x_\nu}{\partial \sigma_2}-
\frac{\partial x_\mu}{\partial \sigma_2}\frac{\partial x_\nu}{\partial \sigma_1}\right) \delta^{(4)}(x-x(\sigma_1,\sigma_2))\nonumber \\
&=&\frac{2\pi}{g} \oint d^2 \sigma_{\mu \nu}\, \delta^{(4)}(x-x(\sigma_1,\sigma_2)),
\end{eqnarray}
where $x(\sigma_1,\sigma_2)$ is the closed two-dimensional surface $\Sigma$ where the singularity is concentrated. In this case, as the surface is closed, proceeding as in eq. (\ref{divd}), we obtain,
\begin{equation}
\partial_\nu d_{\mu \nu}^{(v)} =0.
\end{equation}

\subsection{Correlated defects}

As mentioned in the introduction, according to lattice studies, the relevant configurations could be in fact correlated center vortices and monopoles. It is easy to see that this situation can also be accomodated by using Cho decomposition. In the example at the end of section \S \ref{m-cho}, when discussing a typical monopole configuration with $\alpha=\varphi$, $\beta=\theta$, we considered the case $\gamma =-\varphi$. On the other hand, in that example, a parametrization of the pure monopole sector with $\gamma=0$ cannot be done as this would imply a nonsingle valued mapping. However, as we have seen in section 
\S \ref{center} the center vortex sector can be parametrized on top of the monopole sector $U$ by means of a nonsingle valued $V$. The consideration of $\gamma_v=+\varphi$  in eq. (\ref{Vm}), and the monopole defined by $\gamma =-\varphi$, leads to a well defined local frame to decompose the gauge fields. According to eqs. (\ref{h-ten}) and (\ref{ges}), the Yang-Mills action can be written in terms of $C^{(n)}_\mu$, which in this case is given by,
\begin{equation}
C^{(n)}_\mu=\frac{1}{g} \cos \theta \,  \partial_\mu \varphi,
\label{Cgen}
\end{equation}
which represents two center vortices (on the positive and negative $z$-axis, respectively) attached to a monopole placed at the origin of coordinates.

General configurations with center vortices forming monopole/anti-monopole chains can be similarly pa\-ra\-metri\-zed. Of course, when two center vortices coincide we have in fact a Dirac worldsheet  that is not physical, and that could be changed by means of a gauge transformation.

As correlated monopoles and center vortices can be constructed in terms of a local frame parametrized by the usual $U$ and $V$ sectors, their contribution to $d^{(n)}_{\mu \nu}$ is similar to the above calculation, with the difference that now it is concentrated on  vortex worldsheets attached to monopole worldlines.

\section{Effective model}
\label{emodel}

If closed center vortices are thick, we can consider in eq. (\ref{ZYMb}) the replacement $d^{(n)}_{\mu\nu}=d^{(m)}_{\mu\nu}+d^{(v)}_{\mu\nu}\rightarrow d^{(m)}_{\mu\nu}$, as $d^{(v)}_{\mu\nu}$ is concentrated on $\bar{M}$,
\begin{eqnarray}
\int_M d^4x\, \frac{1}{2} \lambda_{\mu \nu}d^{(n)}_{\mu \nu}=\int_M d^4x\, \frac{1}{2} \lambda_{\mu \nu} d^{(m)}_{\mu \nu}.
\label{mDant}
\end{eqnarray}
We also note that the integration over $A^{(n)}_\mu$, a field living on $M$,
represents a Dirac delta functional $\delta_M[\frac{1}{2}\epsilon_{\mu \nu \rho \sigma} \partial_\nu \lambda_{\rho \sigma}-J^c_\mu]$, defined as a constraint on $M$, which depends on fields not transformed when a gauge transformation is performed (see eqs. (\ref{nonabe-abe}) and (\ref{Jlambda})). 
In addition, $k_{\mu \nu}$ and the last term in the exponent of eq. (\ref{ZYMb}), defined at $\partial M$, are also gauge invariant (cf. eqs. (\ref{felek}) and (\ref{AmasC})). Therefore, if a gauge transformation along the local color direction $\hat{n}$ is considered, the only change in eq. (\ref{ZYMb}) occurs in $d^{(m)}_{\mu \nu}$, the Dirac worldsheet coupled with $\lambda_{\mu \nu}$, leaving the monopoles, represented by $\partial_\nu d^{(m)}_{\mu \nu}=\partial_\nu h^{(m)}_{\mu \nu}$, fixed.

If gauge transformations remain unbroken by infrared quantum effects, the Dirac worldsheets are unobservable, and any choice for them is equally acceptable, only their borders, where monopoles are localized, are physically relevant.
In ref. \cite{unob} we show that when working on the whole Euclidean spacetime, for a given monopole configuration, and for each $B_{\mu \nu}$ realization, it is always possible to choose a Dirac worldsheet such that,
\begin{equation}
\int d^4x\, \frac{1}{2} B_{\mu \nu}d^{(m)}_{\mu \nu}=\frac{2\pi}{g} \int d^2 \sigma_{\mu \nu}\, B_{\mu \nu}=0.
\label{gauge-con}
\end{equation}
The argument to obtain this condition depends on the possibility of continuously deforming 
worldsheets with fixed boundaries one into another. As this is not generally valid on ${\bar M}$, in order to give a simplifying choice, we can consider the extension of the fields inside ${\bar M}$ and write,
\begin{eqnarray}
\lefteqn{\int_M d^4x\, \frac{1}{2} \lambda_{\mu \nu}d^{(m)}_{\mu \nu}}\nonumber \\
&&=\int d^4x\, \frac{1}{2} \lambda_{\mu \nu}d^{(m)}_{\mu \nu}-\int_{\bar{M}} d^4x\, \frac{1}{2} \lambda_{\mu \nu}d^{(m)}_{\mu \nu}\nonumber \\
&&=\int d^4x\, \phi_\mu \partial_\nu d^{(m)}_{\mu \nu}+\int d^4x\, \frac{1}{2} B_{\mu \nu}d^{(m)}_{\mu \nu}-\int_{\bar{M}} d^4x\, \frac{1}{2} \lambda_{\mu \nu}d^{(m)}_{\mu \nu}.
\label{mD}
\end{eqnarray}
On the other hand, using Gauss theorem, the boundary term depending on $C^{(n)}_\mu$ in eq. (\ref{ZYMb}) will give a relevant term, as $C^{(n)}_\mu$ is topologically nontrivial there. At large distances, this term can be estimated by considering the extensions of $\lambda_{\mu \nu}$ and $C^{(n)}_\mu$ inside $\bar{M}$, and using Gauss theorem to write,
\begin{eqnarray}
\lefteqn{\int_M d^4x\, \partial_\mu \left[\frac{1}{2}C^{(n)}_\sigma\epsilon_{\mu \nu \rho \sigma} \lambda_{\nu \rho}\right]=
-\int_{\bar{M}} d^4x\, \partial_\mu \left[\frac{1}{2}C^{(n)}_\sigma\epsilon_{\mu \nu \rho \sigma} \lambda_{\nu \rho}\right]}\nonumber \\
&&=\int_{\bar{M}} d^4x\, \frac{1}{2}\lambda_{\mu \nu}(d^{(v)}_{\mu \nu}+d^{(m)}_{\mu \nu}-h^{(m)}_{\mu \nu})-\int_{\bar{M}} d^4x\, \frac{1}{2}\epsilon_{\mu \nu \rho \sigma} \partial_\mu \lambda_{\nu \rho}\, C^{(n)}_\sigma,
\label{Cboun}
\end{eqnarray}
where the parentesis corresponds to $-\tilde{h}^{(n)}_{\mu \nu}$, rewritten by using eqs. (\ref{hesh}) and (\ref{mv-sep}). Here, as $d^{(v)}_{\mu \nu}$ is concentrated on ${\bar M}$, we can replace $\bar{M}\rightarrow {\cal R}^4$ in the corresponding integral. 
Then, using eqs. (\ref{mDant})-(\ref{Cboun}), we get,
\begin{eqnarray}
\lefteqn{\int_M d^4x\, \frac{1}{2} \lambda_{\mu \nu}d^{(n)}_{\mu \nu}+\int_M d^4x\, \partial_\mu \left[\frac{1}{2}C^{(n)}_\sigma\epsilon_{\mu \nu \rho \sigma} \lambda_{\nu \rho}\right]=}\nonumber \\
&& =\int d^4x\, \phi_\mu \partial_\nu d^{(m)}_{\mu \nu}+\int d^4x\, \frac{1}{2}\lambda_{\mu \nu} d^{(v)}_{\mu \nu}+
{\cal O}(\delta)\nonumber \\
&&=\frac{\pi}{g}\sum_{v} \oint d^2 \sigma_{\mu \nu}\, B_{\mu \nu}+\frac{4\pi}{g}\sum_{ij} \left( \oint_{C^+_j} dy_\mu\, \phi_\mu- \oint_{C^-_i} dy_\mu\,\phi_\mu \right)+{\cal O}(\delta).
\end{eqnarray}
This equation gives the coupling of $\phi_\mu$ with the monopole/anti-monopole ensemble plus the coupling of center vortices and the sector $B_{\mu \nu}$, which represent the U(1) color current $J_\mu^c$.

Terms only depending on regular fields, such as $\lambda_{\mu \nu}$, $A^{(n)}_\mu$, $\Phi_\mu$, when extended to the whole Euclidean spacetime, will introduce additional terms of order $\delta$. Now, when $\delta$ is very small, or equivalently at large distances, we will disregard the terms of order $\delta$. In addition, the contribution associated with the region $\bar{M}$, inside the vortex cores, will be taken into account by simply replacing $S_{\bar M}$ by the large distance behavior $S_\Sigma$ for the physical center vortex profile, as this is the dominant term inside the core. Here, $\Sigma$ is the two-dimensional worldsheet around which thick center vortices are localized ($\Sigma \subset {\bar M}$). The action $S_\Sigma$ is expected to contain a Nambu-Goto term, representing physical objects (see \cite{engelhardt1}), plus other possible terms associated with rigidity \cite{BPZ}. 

In this manner, in order to study the feedback, on gauge fields, of a phase where center vortices become physical and Dirac worldsheets remain unobservable, it is sensible to consider the approximation,
\begin{equation}
Z_{YM}  \approx \int [{\cal D}B][{\cal D}\phi]F^{B}_{gf}F^{\phi}_{gf}\, e^{-S_D[\lambda_{\mu \nu}]-S_{v,m}[B_{\mu \nu},\phi_\mu]},
\label{ZYM-en}
\end{equation}
where $\lambda_{\mu \nu}$ is given by eq. (\ref{lambda}) and,
\begin{eqnarray}
e^{-S_D[\lambda_{\mu \nu}]}&=&
\int [{\cal D}A^{(n)}][{\cal D}\Phi][{\cal D}\Lambda] F^{c}_{gf}\, \times\nonumber \\
&&\times e^{-S_c+i\int d^4x\, [(\frac{1}{2}\epsilon_{\mu \nu \rho \sigma}\partial_\nu \lambda_{\rho \sigma}-J_c^\mu)A^{(n)}_\mu+
\frac{1}{2} \lambda_{\mu \nu} k_{\mu \nu}]}\, \times e^{-\int d^4x\, \frac{1}{4}\lambda_{\mu \nu}\lambda_{\mu \nu}},
\label{Sdual}
\end{eqnarray}
\begin{equation}
e^{-S_{v,m}[B_{\mu \nu}, \phi_{\mu}]}=\int[{\cal D}{\rm vor}][{\cal D}{\rm mon}]\, e^{i\frac{\pi}{g}\sum_{v} \oint d^2 \sigma_{\mu \nu}\, B_{\mu \nu}+i\frac{4\pi}{g}\sum_{ij} \left( \oint_{C^+_j} dy_\mu\, \phi_\mu- \oint_{C^-_i} dy_\mu\,\phi_\mu \right)},
\label{vort-act-corr}
\end{equation}
\begin{equation}
[{\cal D}{\rm vor}]=[{\cal D}V]\, e^{-S_{\Sigma}}.
\end{equation}
If monopoles are uncorrelated with closed center vortices, based on eq. (\ref{vort-act-corr}) we can write,
\begin{equation}
S_{v,m}[B_{\mu \nu}, \phi_{\mu}]=S_v[B_{\mu \nu}]+S_m[\phi_\mu],
\end{equation}
where,
\begin{equation}
e^{-S_v[B_{\mu \nu}]}=\int[{\cal D}{\rm vor}]\, e^{i\frac{\pi}{g}\sum_{v} \oint d^2 \sigma_{\mu \nu}\, B_{\mu \nu}},
\label{vort-act}
\end{equation}
\begin{equation}
e^{-S_m[\phi_{\mu}]}=\int[{\cal D}{\rm mon}]\, e^{i\frac{4\pi}{g}\sum_{ij} \left( \oint_{C^+_j} dy_\mu\, \phi_\mu- \oint_{C^-_i} dy_\mu\,\phi_\mu \right)}.
\end{equation}
On the other hand, for correlated defects, with center vortices forming chains of monopoles and anti-monopoles, it is easy to see that we have to consider in eq. (\ref{ZYM-en}) the replacement $S_{v,m} \to S'_{v,m}$,
\begin{equation}
e^{-S'_{v,m}[B_{\mu \nu}, \phi_{\mu}]}=\int[{\cal D}{\rm vor}][{\cal D}{\rm mon}]\, e^{i\frac{\pi}{g}\sum_{v} \int d^2 \sigma_{\mu \nu}\, B_{\mu \nu}+i\frac{4\pi}{g}\sum_{ij} \left( \oint_{C^+_j} dy_\mu\, \phi_\mu- \oint_{C^-_i} dy_\mu\,\phi_\mu \right)},
\label{vort-act-chains}
\end{equation}
where $B_{\mu \nu}$ is now integrated over vortex worldsheets attached (in pairs) to the corresponding monopole worldlines.

In general, the main nonperturbative information induced by the gluon fluctuations has been included in the modified measure for the ensembles of defects. Similar representations have been discussed in 3D gauge models with monopoles and charged fields (see refs. \cite{CL1,CL2}).

Note that in eq. (\ref{Sdual}), $S_D[\lambda_{\mu \nu}]$ contains no reference to the local frame $\hat{n}_i$, as the fields $A^{(n)}_\mu$, $\Phi_\mu$, ..., are simply dummy variables. Then, this term coincides with the dual action for the linearized form of Yang-Mills theory, discussed in refs. \cite{halpern}-\cite{kondo0}, for the usual gauge field decomposition with respect to the canonical frame. The universal character of the mapping between a charge current and a topological current (cf. eq. (\ref{j-map})) in this type of dual representation has been discussed in ref. \cite{dual-univ}. 

The $\Lambda_{\mu \nu}$ path-integration in eq. (\ref{Sdual}) leads back to the standard quadratic term for the action of a charged field $\Phi_\mu$ coupled with the U(1) gauge field $A^{(n)}_\mu$. As discussed in refs. \cite{halpern}-\cite{kondo0}, as $k_{\mu \nu}$ is also quadratic in the charged fields, the $\Phi_\mu$ path integral gives a functional determinant. The one-loop calculation, including ghosts and lagrange multipliers, has been carried out in ref. \cite{kondo0}. Using that result, the dual action $S_D[\lambda_{\mu \nu}]$ can be written as a gaussian path integral in $A^{(n)}_\mu$, where the exponent contains a combination of 
$f^{(n)}_{\mu \nu}f^{(n)}_{\mu \nu}$, $\epsilon_{\mu \nu \rho \sigma}\partial_\nu \lambda_{\rho \sigma}A^{(n)}_\mu$ and $\lambda_{\mu \nu}\lambda_{\mu \nu}$, which can be performed, with the gauge fixing condition $\partial_\mu A^{(n)}_\mu=0$, to obtain,
\begin{equation}
S_D[\lambda_{\mu \nu}]\approx \int d^4x\, \left[\frac{\gamma}{4} \tilde{H}_\mu \frac{1}{(-\partial^2)} \tilde{H}_\mu +\frac{1}{4}(1+\beta) \lambda_{\mu \nu}\lambda_{\mu \nu}\right],
\label{Sd-approx}
\end{equation}
\begin{equation}
\tilde{H}_\mu=\epsilon_{\mu \nu \rho \sigma}\partial_\nu \lambda_{\rho \sigma}=\epsilon_{\mu \nu \rho \sigma}\partial_\nu B_{\rho \sigma}.
\end{equation}
Integrating by parts the first term in eq. (\ref{Sd-approx}) and the crossed term that appears when using eq. (\ref{lambda}), the dual action can be written in the form,
\begin{equation}
S_D[\lambda_{\mu \nu}]\approx \int d^4x\, \left[\frac{1}{4}(1+\alpha)\, B_{\mu\nu} B_{\mu\nu} +\frac{1}{4}(1+\beta) (\partial_\mu \phi_\nu-\partial_\nu \phi_\mu)^2\right],
\end{equation}
where $\alpha=\gamma+\beta$ and an additional term proportional to $\partial_\nu B_{\mu \nu}$ has been eliminated by taken into account the gauge fixing condition (\ref{g-fixing}), or more precisely, by translating the lagrange multiplier $\xi_\mu$ in eq. (\ref{xi-fixing}) to cancel the additional term. 

For instance, let us consider a phase where closed center vortices, uncorrelated with monopoles, become thick tensile objects. In this case, we can write,
\begin{eqnarray}
\lefteqn{Z_{YM}  \approx \int [{\cal D}B][{\cal D}\phi]F^{B}_{gf}F^{\phi}_{gf}\,e^{-I[\lambda_{\mu \nu}]}\times} \nonumber \\
&& \times e^{-\int d^4x\, \left[\frac{1}{4}(1+\alpha)\, B_{\mu\nu} B_{\mu\nu} +\frac{1}{4}(1+\beta) (\partial_\mu \phi_\nu-\partial_\nu \phi_\mu)^2\right]-S_v[B_{\mu \nu}]-S_m[\phi_\mu]},\nonumber \\
\label{ZYM-eff}
\end{eqnarray}
where $I[\lambda_{\mu \nu}]$ contains perturbative interactions between $B_{\mu \nu}$ and $\phi_\mu$. 

The representation (\ref{ZYM-eff}) is suitable to study the interplay between the different sectors of Yang-Mills theory.  First, we note that $S_v[B_{\mu \nu}]$ in eq. (\ref{vort-act}) is symmetric under the transformation $B_{\mu\nu}+\partial_\mu \chi_\nu -\partial_\nu \chi_\mu$, as the vortices are associated with closed two-dimensional surfaces. 

At large distances, and noting that the mass dimension of $B_{\mu \nu}$ is two, the following form is expected,
\begin{eqnarray}
S_v[B_{\mu \nu}]&\approx &\int d^4x\,\frac{1}{\Lambda_o^2} H_{\mu \nu \rho}H^{\mu \nu \rho}\nonumber \\
&\approx & \int d^4x\,\frac{1}{\Lambda_o^2}\tilde{H}_\mu \tilde{H}_\mu,
\label{ld-v}
\end{eqnarray}
\begin{equation}
\tilde{H}_\mu=\epsilon_{\mu \nu \rho \sigma}\partial_\nu B_{\rho \sigma}
\makebox[.5in]{,}
H_{\mu \nu \rho}=\partial_\mu B_{\nu \rho}+
\partial_\rho B_{\mu \nu}+\partial_\nu B_{\rho \mu}.
\label{HBdual}
\end{equation}

In fact, this can be compared with the situation in 3D Euclidean spacetime. In that case, thick tensile center vortices would be closed strings and $S_{\bar M}$ would contain a term proportional to the string length, with a positive mass (there is an energy cost to enlarge the center vortex). From polymer physics, it is well known that an ensemble of such observable strings can be considered as a second quantized field theory for a complex scalar field (with positive mass) coupled with a field $B_\mu$, which is analogous to $B_{\mu \nu}$, now representing the charged sector in 3D according to,
\begin{equation}
J_c^\mu=\tilde{H}_\mu=\epsilon_{\mu \nu \rho}\partial_\nu B_\rho.
\end{equation}
Because of the mass scale, the large distance effective action after path integrating over the scalar field is naturally dominated by a Maxwell term, $(1/\Lambda_o^2)\, \tilde{H}_\mu \tilde{H}_\mu$.

Similarly, coming back to 4D, for thick tensile center vortices the mass scale in $S_{\bar M}$ is expected to imply the large distance behavior in eqs. (\ref{ld-v}), (\ref{HBdual}) (see also refs. \cite{ambjorn}-\cite{franz}). This, together with the $B_{\mu \nu}^2$ term in eq. (\ref{ZYM-eff}), implies that the charged sector represented by $B_{\mu \nu}$ contains a massive Kalb-Ramond action. 
On the other hand, if no mass scale were associated with center vortices, $S_v[B_{\mu \nu}]$ would be a complicated nonlocal action with no well defined large distance expansion. With regard to the relationship between Yang-Mills theories and a (confining) string theory see ref. \cite{polyakov1,polyakov2}. 

Then, if the off-diagonal mass $\Lambda_o$ is the larger one in the model, greater than the mass scale that can be generated in the monopole sector, the $B_{\mu \nu}$ field is decoupled at large distances, 
\begin{equation}
B_{\mu \nu}\approx 0.
\end{equation}
According to eq. (\ref{j-map}), this can be interpreted by stating that the sector of physical center vortices leads to a phase where a chromoelectric current is driven to zero, decoupling the charged sector in Yang-Mills theories.

In this scenario, the partition function is dominated by a compact QED(4) model for the field $\phi_\mu$. As is well known, this type of model is confining in a phase where monopoles condense \cite{polya1}. As usual, this phase can be analyzed using tools from polymer physics, where the ensemble integration over string-like monopoles, can be represented by means of a second quantized complex field $\psi$, coupled to the gauge field $\phi_\mu$ (see refs. \cite{bardakci}-\cite{halpern2}, \cite{antonov} and references therein). In fact, in ref. \cite{cho5}, the Yang-Mills effective theory in a monopole background has been considered, showing that these objects are unstable. This corresponds to considering the squared mass for $\psi$ as negative, and taking into account contact interactions between the string-like monopoles, a quartic term $\lambda (\bar{\psi}\psi)^2$ in $\psi$-language, to stabilize the system. In other words, the field $\psi$ undergoes a spontaneous symmetry breaking, representing the condensation of monopole degrees of freedom, and $\phi_\mu$ becomes a massive vector field, with a mass scale that has been assumed to be smaller than $\Lambda_0$, the off-diagonal mass that first decouples the charged sector.

\section{Conclusions}
\label{conc}

Recent studies in pure Yang-Mills theory, that include as nonperturbative information the Gribov horizon and possible condensates, point to infrared suppressed gluon and ghost propagators. If on the one hand this behavior implies the abscence of gluons in the asymptotic spectrum of the theory, it raises the problem of how  long range confining quark interactions can be implied. One promising source for this type of behavior resides in the inclusion of topological defects, looking at the possible phases that could be induced by nonperturbative gluon fluctuations. This is one of the reasons to search for a simple description of continuum Yang-Mills theory including all topological sectors. In the first part of this work, we have obtained such a description, by unifying monopoles and center vortices as different classes of defects in the local color frame $\hat{n}_a=R\, \hat{e}_a$, $R\in$ SO(3), used in Cho decomposition.

In the second part, relying on this procedure, where center vortices are constructed on top of monopole configurations, we have derived an effective model for Yang-Mills theory where the main nonperturbative information has been parametrized in a modified measure for the ensembles of defects. 
In this scenario, abelian dominance can be understood as the feedback,  on the charged modes, of a phase where center vortices become physical, leaving a compact abelian theory which is expected to be confining in a phase where monopoles condense.

\vspace{.15in}

When studying abelian projection scenarios, the gauge fields are generally separated into ``diagonal'' fields, living in the Cartan subalgebra of SU(N), and ``off-diagonal'' charged fields. 
For instance, in the case of SU(2), the uncharged sector can be chosen along the $\hat{e}_3$ direction in color space, while the components along $\hat{e}_1$ and $\hat{e}_2$ correspond to the charged sector.  
In Cho decomposition, this separation into charged and uncharged sectors, with respect to an abelian subgroup of SU(2) rotations, is also implemented,  with the advantage that it is naturally done along a general $\hat{n}_3\equiv \hat{n}$ local direction in color space. 

Compared with the field strength tensor computed in refs. \cite{cho2}-\cite{cho5}, we have identified two types of singular terms, when working with color frames containing defects. The first one, $\vec{L}_{\mu \nu}$ in eq. (\ref{FL}), depends on defects of the third component $\hat{n}_3\equiv \hat{n}$, and occurs in the charged sector of the field strength tensor. In section \S \ref{YMch}, we have seen that in order to preserve the U(1) symmetry of the theory, associated with phase transformations of the charged field $\Phi_\rho=\frac{1}{\sqrt{2}}(X^1_\rho+iX^2_\rho)$, 
the possible singularities are restricted by the condition $\vec{L}_{\mu \nu}=\vec{0}$. This implies that $\hat{n}$ can have at most defects concentrated on closed strings. As is well known, this corresponds to monopoles, characterized by $\Pi_2(S^2)=Z$, and a  piece $h^{(n)}_{\mu \nu}$, with nontrivial divergence, added to the the dual field strength $f^{(n)}_{\mu \nu}$, in the zero charge sector of the theory.

The second type occurs when trying to express the monopole part $h^{(n)}_{\mu \nu}$ of the dual field strength 
in terms of the monopole potential $C_\mu$. When the components $\hat{n}_1$ and $\hat{n}_2$ contain defects, $h^{(n)}_{\mu \nu}$ and $\tilde{h}^{(n)}_{\mu \nu}=\epsilon_{\mu \nu \rho \sigma}\partial_\rho C_\sigma$ differ by singular terms which are important to understand the possible consequences of the ensembles of defects in Yang-Mills theory. 

In particular, we have seen that when a monopole singularity for $\hat{n}$ occurs, this is accompanied by defects of $\hat{n}_1$ and $\hat{n}_2$ on an open two-dimensional worldsheet, having the monopoles at the boundaries. This defect corresponds to the Dirac worldsheet. If we stay close to the worldsheet and go around it once, the components $\hat{n}_1$, $\hat{n}_2$ rotate twice. This is associated with the magnetic flux $4\pi/g$ carried by the Dirac worldsheet, matching the magnetic flux $4\pi/g$ emanating from monopoles in nonabelian theories. For fixed monopole positions, the Dirac worldsheet can be changed by performing a topologically trivial SU(2) gauge transformation, representing a rotation along the $\hat{n}$ direction. As long as this symmetry is not broken, the Dirac worldsheets remain unobservable objects.

The discussion above led us to consider other possible defects of the local color frame, concentrated on closed two-dimensional surfaces, such that the components $\hat{n}_1$, $\hat{n}_2$ rotate once, when we go around the defect, and the third component $\hat{n}$ is nonsingular on these surfaces. Precisely, these defects correspond to center vortices carrying magnetic flux $2\pi/g$.  In short, center vortices can be associated with the nontrivial first homotopy group $\Pi_1=Z(2)$ of SO(3), which defines a general local color frame by $\hat{n}_a=R\, \hat{e}_a$, $R\in $ SO(3). A general distribution of defects can be obtained by constructing center vortices on top of the monopole configurations.

The important point is that although $R\in$ SO(3) can be thought of as an adjoint representation of an SU(2) mapping, when center vortices are present, they cannot be associated with a topologically nontrivial ``gauge'' transformation. In this case, the necessary SU(2) transformation would not be single valued, and additional defects localized on the three-volumes where this transformation is discontinuous would also be introduced (ideal vortices).

This opens the possibility of two different behaviors. While open Dirac worldsheets remain unobservable, no symmetry is present to protect thin center vortices from a destabilization into physical thick center vortices. 

In fact, the use of center vortices and monopoles as a reasonable phase, that serves as a background to include gluon fluctuations, depends on their stability. In turn, this stability can be studied by means of the path integration over gluon fields, analyzing if the probability to persist in the fundamental state, in the given background, is less than one.
Indeed, lattice simulations point to the idea that center vortices become stabilized by generating a finite radius, with a thickness of the order of 1fm. In that case, the parametrization of center vortices in terms of defects of a local frame is only valid outside the center vortex cores.

After gauging out the Dirac worldsheets, we obtained a representation for the partition function where the main nonperturbative information induced by gluon fluctuations has been pa\-ram\-e\-trized in a modified measure for general ensembles of center vortices and monopoles, which are coupled with the dual form of SU(2) Yang-Mills theory. This dual form depends on fields $\phi_\mu$, $B_{\mu \nu}$, and coincides with the one that would be obtained in Yang-Mills theory without defects, and using the usual canonical basis to decompose the gauge fields. 

If center vortices were thin objects, the associated ensemble integration would imply complicated nonlocal terms for $B_{\mu \nu}$, with no well defined large distance expansion. By the way, in this case, a representation similar to the one presented here could be useful to study the possible effect of nonperturbative gluon fluctuations on monopoles and thin center vortices.

On the other hand, in a phase where center vortices become thick tensile objects, uncorrelated with monopoles, we have shown that a massive Kalb-Ramond term is expected to be induced. If this ``off-diagonal'' mass happens to be the larger one in the model, the $B_{\mu \nu}$ field decouples at large distances, $B_{\mu \nu}\approx 0$, and the partition function is dominated by a compact QED(4) model, which is known to be confining when monopoles condense (see \cite{polya,polya1}, \cite{Panero} and references therein). 

The scenario we have presented can be compared with other destabilizations in Field Theory. For instance, if a Fermi liquid in 1D, associated with gapless fermion modes in (1+1)D, is coupled to phonons, the path integral over the gapless fermions induces a Peierls instability, where the phonon field aquires a nonzero expectation value. In turn, the effect of this instability can be analyzed by coupling, from the beginning, the previously gapless fermion modes with the modified phase for the phonon field. This feedback leads to fermion modes that acquire a gap, that is, a metal-insulator transition is induced. 

In our case, from a physical point of view, because of the dual mapping $J^c_\mu=(1/2)\epsilon_{\mu \nu \rho \sigma} \partial_\nu B_{\rho \sigma}$,  the decoupling of $B_{\mu \nu}$ can be interpreted by stating that the sector of physical center vortices leads to a phase where a chromoelectric current is driven to zero, thus characterizing a superconductor of chromomagnetic charges. This also suggests that the feedback of an induced finite radius for center vortices, due to gluon fluctuations, is the decoupling of the charged sector in Yang-Mills theories, that is, abelian dominance.

It is important to emphasize that how thin center vortices could be destabilized into a phase of thick vortices is still an open problem. We believe that it would be interesting to pursue further studies to characterize the stable phase when monopoles and center vortices coexist.

In this regard, we would like to underline the following points. 

Following Petrov-Diakonov approach \cite{PD}, the procedure we have followed here for the partition function can be adapted to represent the Wilson loop and discuss the conditions to obtain an area law and the associated long range confining interactions \cite{B}. As occurs in eqs. (\ref{vort-act-corr}) and (\ref{vort-act-chains}), where the monopole sector couples with $\phi_\mu$ and the center vortex sector couples with $B_{\mu \nu}$, in the Wilson loop average there will be terms where monopoles and center vortices couple with $\psi_{\mu}$ and $R_{\mu \nu}$, respectively; these fields are defined similarly to eq. (\ref{lambda}) by,
\begin{equation}
s_{\mu \nu}=\partial_\mu \psi_\nu-\partial_\nu \psi_\mu + R_{\mu \nu},
\label{l-s}
\end{equation}
with,
\begin{equation}
\partial_\mu \psi_\mu=0  \makebox[.5in]{,}  \partial_\nu R_{\mu \nu}=0,
\label{g-fixing-bp}
\end{equation}
where $s_{\mu \nu}$ is a source with support on a surface having the Wilson loop as boundary. That is,  $\psi_\nu$ depends on $\partial_\mu s_{\mu \nu}$, concentrated on the surface, while $R_{\mu \nu}$ depends on $\epsilon_{\mu \nu \rho \sigma}\partial_\nu s_{\rho \sigma}$, concentrated on the perimeter.

In the above mentioned phase where closed center vortices become thick tensile objects, and they are uncorrelated with monopoles, the vortex integration will lead at large distances to new terms depending on the perimeter, so that in this case the area law is expected to be only due to the monopole sector. Note that this situation corresponds to nonpercolating vortices, as the large distance behavior in eq. (\ref{ld-v}) corresponds to an ensemble where large tensile objects have vanishing weight.

On the other hand, when center vortices percolate, the large distance behavior in eq. (\ref{ld-v}) cannot be applied. In this case, the ensemble integration in eqs. (\ref{vort-act-corr}) and (\ref{vort-act-chains}) has to be discussed separately. 

In this respect, it is important to note that the coupling of closed center vortices with $R_{\mu \nu}$ gives the linking between them and the Wilson loop. In the lattice, when center vortices percolate, the average of $(-1)^n$, where $n$ is the number of vortices linking the loop, is known to be associated with a confining phase (see \cite{greensite} and references therein). It is also interesting to note that this type of factor would also be present in our representation when chains of monopoles and anti-monopoles joined by center vortices are considered (see ref. \cite{B}). In this case, $n$ would be given by the number of chains linking the Wilson loop. However, differences when compared with a model only containing percolating center vortices are expected, because of the new couplings between defects and the dual fields $\phi_\mu$ and $B_{\mu \nu}$.

Here, we have shown that the role of a phase of tensile center vortices is that of abelianizing the theory at large distances. In this case, the area law is produced by an uncorrelated sector of monopoles. However, it would be very interesting to study a phase formed by percolating chains of monopoles and anti-monopoles as according to refs. \cite{AGG}-\cite{GKPSZ} they could be the relevant configurations to accomodate all the properties that should be displayed by the confining potential between quarks. The representation we have derived could play an important role to implement such studies in the continuum.

\section{Acknowledgements}
I am indebted to S. P. Sorella and M. Moriconi for useful discussions. The Conselho Nacional de Desenvolvimento Cient\'{\i}fico e Tecnol\'{o}gico (CNPq-Brazil), the Funda{\c {c}}{\~{a}}o de Amparo {\`{a}} Pesquisa do Estado do Rio de Janeiro (FAPERJ), and the Pr\'o-Reitoria de P\'os-Gradua\c c\~ao e Pesquisa da Universidade Federal Fluminense (PROPP-UFF), are acknowledged for the financial support.

\appendix

\section{Covariant derivative in ``abelianized'' form}

Using eqs. (\ref{Arest}) and (\ref{Drest}), we obtain,
\begin{eqnarray}
\lefteqn{\hat{D}_\mu \vec{X}^{(n)}_\nu= \hat{n}_1(\partial_\mu X^1_\nu-gA^{(n)}_\mu X^2_\nu)+\hat{n}_2(\partial_\mu X^2_\nu+ gA^{(n)}_\mu X^1_\nu)}\nonumber \\
&& +X^1_\nu\partial_\mu \hat{n}_1 +X^2_\nu\partial_\mu \hat{n}_2 ++\hat{n} [(X^1_\nu \hat{n}_1+X^2_\nu \hat{n}_2).\partial_\mu \hat{n}].
\end{eqnarray}
Now, projecting with $\hat{n}_i$, we get,
\begin{equation}
\hat{n}_1.\hat{D}_\mu \vec{X}^{(n)}_\nu=\partial_\mu X^1_\nu-gA^{(n)}_\mu X^2_\nu +\hat{n}_1.\partial_\mu \hat{n}_2 X^2_\nu,
\end{equation} 
\begin{eqnarray}
\hat{n}_2.\hat{D}_\mu \vec{X}^{(n)}_\nu&=&\partial_\mu X^2_\nu+gA^{(n)}_\mu X^1_\nu +\hat{n}_2.\partial_\mu \hat{n}_1 X^1_\nu\nonumber \\
&=&\partial_\mu X^2_\nu+gA_\mu X^1_\nu -\hat{n}_1.\partial_\mu \hat{n}_2 X^1_\nu,
\end{eqnarray}
\begin{eqnarray}
\hat{n}.\hat{D}_\mu \vec{X}^{(n)}_\nu&=&= X^1_\nu \hat{n}.\partial_\mu \hat{n}_1
+X^2_\nu\hat{n}.\partial_\mu \hat{n}_2 +\nonumber \\
&& +(X^1_\nu \hat{n}_1+X^2_\nu \hat{n}_2).\partial_\mu \hat{n}=0,
\end{eqnarray}
and defining,
\begin{equation}
-g C^{(n)}_\mu= \hat{n}_1.\partial_\mu \hat{n}_2,
\end{equation}
we obtain eq. (\ref{DX}).

\section{Decomposition of ``pure gauge'' fields}

For a generally nontrivial $U\in $ SU(2), single valued along any closed loop, we have,
\begin{eqnarray}
\frac{i}{g}U\partial_\mu U^{-1} &=&\frac{2i}{g} \sum_a tr\, (U\partial_\mu U^{-1} \hat{m}_a.\vec{T})\, \hat{m}_a.\vec{T}\nonumber \\
&=&\frac{2i}{g} \sum_a tr\, (U\partial_\mu U^{-1}\, UT^a U^{-1})\, \hat{m}_a.\vec{T}\nonumber \\
\nonumber \\
&=&-\frac{i}{g} \sum_a tr\, ( U^{-1} \partial_\mu U\tau^a)\, \hat{m}_a.\vec{T}.\nonumber \\
\label{three}
\end{eqnarray}
On the other hand, recalling that $\tau^3=2 T^3$, $iT^3=[T^1,T^2]$, we can write,
\begin{eqnarray}
\frac{i}{g} tr\,[ \tau^3 U^{-1}\partial_\mu U]&=&\frac{2}{g}tr\, 
[ T^1 T^2 U^{-1}\partial_\mu U- T^2 T^1 U^{-1}\partial_\mu U]\nonumber \\
&=&-\frac{2}{g}tr\, 
[ T^1 T^2 \partial_\mu U^{-1} U+ T^2 T^1 U^{-1}\partial_\mu U]\nonumber \\
&=&-\frac{2}{g}tr\, 
[(U T^1 U^{-1}) U T^2 \partial_\mu U^{-1} +  (UT^1 U^{-1})\partial_\mu UT^2U^{-1}]\nonumber \\
&=&-\frac{2}{g}tr\, 
[(U T^1 U^{-1}) \partial_\mu (UT^2U^{-1})].
\end{eqnarray}
Then, defining $U T^a U^{-1}=\hat{m}_a.\vec{T}$, and using $tr\, (T^a T^b)=\frac{1}{2}\delta^{ab}$, we get,
\begin{equation}
\frac{i}{g} tr\,[ \tau^3 U^{-1}\partial_\mu U]=-\frac{1}{g}\hat{m}_1.\partial_\mu \hat{m}_2=C^{(m)}_\mu.
\end{equation}
Working in a similar manner with the other two terms in eq. (\ref{three}),
\begin{equation}
\frac{i}{g}U\partial_\mu U^{-1}=(-C^{(m)}_\mu \hat{m} +\frac{1}{g}(\hat{m}_2.\partial_\mu \hat{m})\, \hat{m}_1 +
\frac{1}{g}(\hat{m}.\partial_\mu \hat{m}_1)\, \hat{m}_2).\vec{T}
\label{pg-sym}
\end{equation}
where we have defined $\hat{m}_3=\hat{m}$. Now, noting that,
\begin{eqnarray}
-\frac{1}{g}\hat{m}\times\partial_\mu \hat{m}&=& -\frac{1}{g}\hat{m}\times\partial_\mu (\hat{m}_1\times\hat{m}_2)\nonumber \\
&=& \frac{1}{g}( \hat{m}.\partial_\mu \hat{m}_1)\, \hat{m}_2\,  + (\hat{m}_2.\partial_\mu \hat{m})\, \hat{m}_1,
\end{eqnarray}
we finally obtain,
\begin{equation}
\frac{i}{g} U\partial_\mu U^{-1}=-(C^{(m)}_\mu \hat{m}+\frac{1}{g}\hat{m} \times \partial_\mu \hat{m}).\vec{T}.
\label{U-trans}
\end{equation}
The results above have been previously obtained in \cite{cho-07}, by following a different route. 
Note also that the expression $\frac{i}{g} U\partial_\mu U^{-1}$ in eq. (\ref{pg-sym}) treats the different elements of the frame symmetrically, so that we can single out any of them to make the decomposition.


\end{document}